\documentstyle[aps,epsf,multicol]{revtex} 
 
\def\be{\begin{equation}} 
\def\ee{\end{equation}} 
\def\bea{\begin{eqnarray}} 
\def\eea{\end{eqnarray}} 
 
\newcommand{\e}{{\rm e}} 
\newcommand{\half}{\mbox{$\frac{1}{2}$}} 
\newcommand{\tmu}{{\tilde \mu}} 
\newcommand{\Bt}{{\tilde B}} 
 
\newcommand{\lb}{\label}

\begin{document} 
 
\draft \title{Embedding a Native State into a Random \\ 
Heteropolymer Model: The Dynamic Approach} 
 
\author{Z. Konkoli$^{1,2}$ and J. Hertz$^{2}$} 
\address{ 
  $^1${Department of Applied Physics, \\ 
       Chalmers University of Technology and G\"oteborg University, \\ 
       SE 412 96 G\"oteborg, Sweden}   \\ 
  $^2${NORDITA, Blegdamsvej 17, DK 2100 K\o benhavn, Denmark} 
} 
\date{\today} 
\maketitle 
 
\begin{abstract} 
We study a random heteropolymer model with Langevin dynamics, in the 
supersymmetric formulation.  Employing a procedure similar to one that 
has been used in static calculations, we construct an ensemble in 
which the affinity of the system for a native state is controlled by a 
``selection temperature" $T_0$.  In the limit of high $T_0$, the model 
reduces to a random heteropolymer, while for $T_0 \rightarrow 0$ the 
system is forced into the native state.  Within the Gaussian 
variational approach that we employed previously for the random 
heteropolymer, we explore the phases of the system for high and low 
$T_0$.  For high $T_0$, the system exhibits a (dynamical) spin glass 
phase, like that found for the random heteropolymer, below a 
temperature $T_g$.  For low $T_0$, we find an ordered phase, 
characterized by a nonzero overlap with the native state, below a 
temperature $T_n \propto 1/T_0 > T_g$.  However, the random-globule 
phase remains locally stable below $T_n$, down to the dynamical glass 
transition at $T_g$.  Thus, in this model, folding is rapid for 
temperatures between $T_g$ and $T_n$, but below $T_g$ the system can 
get trapped in conformations uncorrelated with the native state.  At a 
lower temperature, the ordered phase can also undergo a dynamical 
glass transition, splitting into substates separated by large 
barriers. 
\end{abstract} 
 
 
\maketitle 
 
\begin{multicols}{2} 
\narrowtext 
 
\section{Introduction} 
 
The protein folding process is relevant for all aspects of life: 
once read off from the RNA chain, proteins perform a variety of 
functions, from mechanical work to attacking viruses.~\cite{Boy} 
The key factor which determines the function of a protein molecule 
is its 3D structure, which, in turn, is determined by the sequence 
of amino acids forming the protein 
chain.~\cite{PGT1,Wol1,WE,Creig} Furthermore, a protein that has 
been denatured (by stretching it for example) finds its native 
state relatively quickly.  Protein folding has attracted an 
enormous amount of scientific attention, but still there is no 
generic understanding of this process.  Nevertheless, one thing is 
clear: a proteins generally has a potential energy surface which 
results in a stable free energy minimum, corresponding to the 
native state~\cite{Wol1}. 
 
Random heteropolymer models (RHP) have been used extensively as 
candidate systems which might help us to understand the generic 
features of the potential energy surfaces of proteins and their 
connection with thermodynamic 
~\cite{SG1,SG2,GHLO,GOP,GLO,TW,SGS,SW} and dynamical 
~\cite{LT,Pit,TPW,TAB,PS,Olem1,Olem2} properties.  The RHP model 
is characterized by quenched random monomer-monomer interactions, 
meant to mimic the variety of interactions between amino-acids in 
random sequences.  It turns out that the potential energy surface 
of the RHP is quite similar to that of a particular class of spin 
glasses~\cite{SpGl}: Its complex form, with exponentially large 
numbers of local minima and saddle points, constrains the motion 
of the system drastically, and it cannot explore its full 
configuration space and reach Gibbs equilibrium.  In a previous 
paper (\cite{KHS}, henceforth referred to as paper I), we 
demonstrated, in mean field theory, the existence of a sharp 
transition  to a ``dynamical glassy state'' in which the 
equilibration time diverges and the dynamics exhibit aging. (The 
potential importance of spin glass physics to proteins was first 
discussed in Ref.~\cite{Wol2}). Obviously, the random 
heteropolymer model does not describe a protein with a native 
state, but it alerts us to the need to examine possible glassiness 
in models for protein dynamics. 
 
Why are real proteins not glassy?  Evidently, nature has tuned amino 
acid sequences to avoid glassy behavior.  To understand how such 
tuning might be done, it is worthwhile to study models which contain 
competition between glassiness and a tendency to form a native state, 
by choosing interactions which are not completely random.  Several 
studies along the lines of this suggestion have been made in {\em 
statics} (using the replica treatment, see, e.g., Ref.~\cite{Wol2}). 
The tendency toward a particular state can be built in by choosing 
sequences from a distribution correlated with the native sequence 
\cite{PGT1,RS,PGT2,WS}.  A dynamical treatment of similar models is 
highly desirable, not only to help gain insight into results obtained 
in replica approaches, but also because knowledge of the correct 
thermodynamics alone may not be sufficient: it is known that in 
related (mean field models) static and dynamic phase diagrams can be 
different.  Thus (at least on sufficiently short time scales) only a 
dynamical approach can describe the measurable properties of the 
system.  In this paper we undertake such a study. 
 
We extend the RHP model studied in \cite{SG1,SG2} to include the 
existence of a native state: the original random monomer-monomer 
interactions are biased so as to favor the native state 
conformation. The problem is formulated as a Langevin model.  To 
the best of our knowledge, there is so far neither a static nor a 
dynamic treatment available for a model of this sort: Static 
studies have been based on random monomer sequences, i.e., using 
only $N$ random parameters, see Refs.~\cite{PGT1,RS,PGT2,WS}, 
rather than the $N(N-1)/2$ in the RHP model. 
 
Admittedly, the model does not describe a realistic protein (e.g., 
it does not give rise to secondary structure such as 
$\alpha$-helices or $\beta$-sheets). However, it does contain 
important generic features: the polymeric structure and the 
mixture of attractive and repulsive interactions.  Together, these 
features lead to frustration in the structural dynamics.  In our 
view, ours is the simplest such model that includes competition 
between glassy and native states.  As we will see, it teaches us 
that one can not get rid of glassiness so easily. 
 
As in paper I, we simplify the model further by omitting 
three-body interactions in the polymer. (A review describing how 
to include three-body terms is given in~\cite{GOP}.) The price we 
have to pay for this simplification is that we have to introduce a 
somewhat arbitrary confining potential, which we take to have a 
quadratic form.  We adjust its strength so that the radius of 
gyration $R_g$ of a polymer of size $N$ scales like $N^{-1/d}$, 
where $d$ is the dimensionality of the system. In this way we 
attempt to describe a globular state. Of course, we can not 
describe the $\theta$-point transition in such a model, but here 
we are only interested in transitions between different globular 
states. 
 
Our formal starting point is the Martin-Siggia-Rose generating 
functional for the Langevin dynamics of the model 
\cite{MSR,Dom,Jans1,Jans2}, written, for convenience and 
compactness, in its supersymmetric form \cite{Kur}.  To derive 
equations of motion for correlation and response functions we use 
a variational ansatz with a quadratic action. This approach has 
been used to study the problem of a manifold in a random 
potential, in both statics \cite{MP1,MP2} and dynamics 
\cite{CKD,CD}. 
 
In paper I we showed that the RHP model exhibited broken 
ergodicity (formally, a spontaneous supersymmetry breaking) in a 
low-temperature dynamical glassy phase. In the present study, with 
interactions biased in favor of a native state to a controlled 
degree, we find, in addition, a well-folded phase, if the bias is 
strong enough.  It can coexist with either the disordered 
(random-globule) state or the frozen-globule glass phase, 
depending on the temperature. Furthermore, we find that at low 
temperature the native phase can itself undergo a dynamical 
freezing into a different glassy phase. In this phase the 
conformation of the protein is always highly correlated with the 
native state, but cooperative kinetic constraints still lead to a 
divergent equilibration time, as for the frozen-globule state.

\section{The Model} 
 
The model is defined as follows. The Langevin dynamics is assumed to 
be governed by a Hamiltonian $H[x]$, 
\begin{equation} 
  \partial x(s,t)/\partial t = - \delta H[x] / \delta x(s,t) + \eta(s,t). 
  \label{eq:dxdt} 
\end{equation} 
Here $x(s,t)$ is the position of monomer $s$ at time $t$.  The monomers 
are numbered continuously from $s=0$ to $s=N$. $\eta(s,t)$ is Gaussian 
noise 
\begin{equation} 
  \langle \eta(s,t)\eta(s',t') \rangle_T = 2T\delta(s-s')\delta(t-t'), 
  \label{eq:etas} 
\end{equation} 
resulting from coupling to a heat bath at temperature $T$. 
 
The Hamiltonian $H[x]$ contains a deterministic part $H_0[x,\mu]$ 
and a random part $H[x,\{B\}]$.  $H_0[x,\mu]$ is defined as 
\begin{equation} 
   H_0[x,\mu]= \frac{T}{2} \int_{0}^{N} ds 
        \{[\partial x(s,t)/\partial s]^2+\mu x(s,t)^2\}. 
   \label{eq:H0} 
\end{equation} 
It describes the elastic properties of the chain and a confinement 
potential which fixes the density of the protein.  The radius of 
gyration $R_g \sim \mu^{-1/4}$, so, in order that the protein is 
compact, i.e., $R_g \sim N^{1/d}$,  we require $\mu \sim 
N^{-4/d}$. Thus, since we are interested in very long proteins (to 
obtain the thermodynamic limit) we need to solve the model for 
$\mu$ close to zero. 
 
The random part $H[x,\{B\}]$ describes the quenched random 
interactions between monomers, 
\begin{equation} 
   H[x,\{B\}]= \frac{1}{2} \int_{0}^{N} ds ds' 
              B_{ss'} V(x(s,t)-x(s',t)). 
   \label{eq:Hrand} 
\end{equation} 
We take $B_{ss'}$ Gaussian, with variance $B^2$.  The 
quenched average over $B_{ss'}$ is performed as $\langle (.) \rangle_B 
= \int \prod_{s>s'} dB_{ss'} (.) P(\{B\})$. 
$V(\Delta x)$ is a short-range potential, and, for simplicity, we 
take it to have a Gaussian form, as in Ref.~\cite{TPW}, 
\begin{equation} 
   V(\Delta x)=\left(\frac{1}{2\pi\sigma}\right)^{d/2} 
   \e^{-(\Delta x)^2/2\sigma}. 
   \label{eq:V} 
\end{equation} 
$d$ is the dimensionality of the system and $\sqrt{\sigma}$ the range of 
the potential.  Large (small) $\sigma$ corresponds to a long (short) 
range potential. In particular, for $\sigma\rightarrow 0$, $V(\Delta 
x)\rightarrow\delta(\Delta x)$, and we recover the potential used in 
\cite{SG1,SG2,PS}. Here and in the following $\Delta x$ refers to a 
monomer-monomer distance: $\Delta x=x(s,t)-x(s',t)$ for a pair of monomers 
$s$, $s'$. 
 
We use reasoning similar to that employed in statics to define $P(\{B\})$ 
(see Refs.~\cite{PGT1,RS,PGT2,WS}), adapting it to the random-bond model: 
\begin{equation} 
   P(\{B\}) \propto \e^{ - \frac{1}{T_0} H[x_0,{B}] 
                      - \frac{1}{2}\int ds ds' B_{ss'}^2 / 2 B^2 } 
   \label{eq:PB1} 
\end{equation} 
$T_0$ is called the selection temperature, and $x_0(s)$ is some arbitrary 
native state conformation.  Thus the symmetric bond distribution of the 
RHP model is distorted so as to give bigger weight to $B_{ss'}$'s which 
are attractive between monomers which lie close to each other in the 
configuration $x_0(s)$.   Explicitly, the properly normalized 
$P(\{B\})$ is given by 
\begin{eqnarray} 
 && P(\{B\}) = (2\pi B^2)^{-N(N-1)/4} 
 \e^{-\beta_0^2 B^2 /4 \int ds ds' V(x_0(s)-x_0(s'))^2} 
                \nonumber \\ 
 && \times \e^{-\beta_0/2\int dsds'B_{ss'} V(x_0(s)-x_0(s')) 
               -1/2\int dsds'B_{ss'}^2/2B^2}, 
  \label{eq:PB2} 
\end{eqnarray} 
from which we see that the distribution of $B_{ss'}$ is peaked around 
$B^{max}_{ss'} = - \beta_0 B^2 V(x_0(s)-x_0(s'))$. Thus, if monomers 
$s$ and $s'$ are close in the native state ($V(x_0(s)-x_0(s'))\ne 0$), 
their coupling constant $B_{ss'}$ is pulled down, as in a Go model 
\cite{Go1,Go2}.  For $T_0\rightarrow\infty$ we recover the RHP model. 
For $T_0\rightarrow 0$, $P(\{B\})$ picks a specific set of 
$B_{ss'}$. For this set, by construction, $x_0(s)$ is the deepest 
minimum of $H[x,\{B\}]$ given in Eq.~(\ref{eq:Hrand}).  This is the 
mechanism that embeds the native state $x_0(s)$. 
 
This mechanism is somewhat arbitrary. However, the fact that the 
strength of embedding of the native state is controlled by the single 
parameter $T_0$ facilitates the study of transitions between random 
and native-like states (and, as we will show, of possible coexistence 
of such phases). 
 
So far, the configuration $x_0(s)$ is arbitrary.  Thus $x_0(s)$ has to 
be considered a quenched random function, to be averaged over just 
like $B_{ss'}$ in order to obtain generic results.  We will carry this 
average out later. 
 
All our results are obtained in the thermodynamic limit, where the 
length $N$ of the heteropolymer chain goes to infinity.  Also, for 
simplicity, we join the polymer ends to form a ring.  This neglect 
of end effects is valid for a long chain.

\section{Mapping to the Field Theory} 
 
To solve the model we map the Langevin dynamics onto a 
supersymmetric (SUSY) field theory.  Using the standard 
Martin-Siggia-Rose formalism \cite{MSR,Dom,Jans1,Jans2} and 
supersymmetric (SUSY) notation \cite{Olem1,Olem2,Kur,Olem3}, the 
dynamical average of any observable, for fixed $\{B\}$, can be 
calculated as (see, e.g., Paper I for details), 
\begin{eqnarray} 
  && \langle {\cal O}[\Phi] \rangle_T =\int D\Phi 
    {\cal O}[\Phi] e^{-S[\Phi] },           \label{eq:avSUSY} \\ 
  && S[\Phi] = S_1[\Phi]+S[\Phi, x_0, \{B\}],      \label{eq:SSUSY} 
\end{eqnarray} 
where 
\begin{eqnarray} 
  & & S_1[\Phi]  =  1/2 \int ds d1 ds' d2 \Phi(s,1) K_{12}^{ss'} \Phi(s'2), 
  \label{eq:S0} \\ 
  & & S[\Phi,x_0,\{B\}]  =  1/2 \int d1 ds ds' \times \nonumber \\ 
  &   & \ \ \ \ \ \ \ \ \ \ \ \ \ \ \ \ \ \ 
        \ \ \ \ \ \ \times B_{s,s'} V(\Phi(s,1)-\Phi(s',1)), 
  \label{eq:Srand} 
\end{eqnarray} 
and 
\begin{eqnarray} 
  K_{12}^{ss'} && \equiv \delta_{12} \delta_{ss'} K_1^s \ , \ \ 
  K_1^s  = T \left[ \mu-(\partial/\partial s)^2 \right] - D_1^{(2)}, \\ 
  D_1^{(2)} && =2 T \frac{\partial^2}{\partial\theta_1\partial\bar\theta_1} 
  +  2 \theta_1 \frac{\partial^2}{\partial\theta_1\partial t_1} - 
  \frac{\partial}{\partial t_1}. 
\end{eqnarray} 
The $\Phi(s,1)$ denotes a superfield 
\begin{eqnarray} 
  & \Phi(s,1) = & x(s,t_1) + \bar\theta_1 \eta(s,t_1) + \nonumber \\ 
  &             & + \bar\eta(s,t_1) \theta_1 
                  + \bar\theta_1\theta_1\tilde x(s,t_1) 
  \label{Phis1} 
\end{eqnarray} 
containing the physical coordinate $x(s,t)$, the MSR auxiliary field 
$\tilde x(s,t)$, ghost fields $\eta(s,t)$ and $\bar\eta(s,t)$ that 
enforce the normalization of the distribution, and Grassmann variables 
$\theta$ and $\bar\theta$.  We use the notation $1\equiv 
(\theta_1,\bar\theta_1,t_1)$, likewise $\int d1 \equiv \int d\bar\theta_1 
d\theta_1 dt_1$. 
 
Of course, the solution can be obtained without the aid of the 
supersymmetric formalism, but we find it conveniently compact.

As noticed by De Dominicis \cite{Dom} the expression in 
Eq.(\ref{eq:avSUSY}) is already normalized, so the average over the 
quenched disorder $B_{s,s'}$ can be done directly on 
(\ref{eq:avSUSY}): 
\begin{equation} 
  \langle\langle {\cal O}[\Phi] \rangle_T\rangle_B = \int D\Phi 
  {\cal O}[\Phi] \e^{-(S_1[\Phi]+S_2[\Phi, x_0])} 
  \label{avO}, 
\end{equation} 
where $\exp(-S_2[\Phi, 
x_0])\equiv\langle\exp(-S[\Phi,x_0,\{B\}])\rangle_B$, and 
\begin{eqnarray} 
  & S_2[\Phi, x_0] = & -\frac{B^2}{4} \int ds ds' 
                  \left[ 
                      \int d1 V(\Phi(s,1)-\Phi(s',1)) 
                  \right]^2  \nonumber \\ 
  && - \frac{\beta_0 B^2}{2} \int ds ds' d1 V(\Phi(s,1)-\Phi(s',1)) \times 
     \nonumber \\ 
  && \times V(x_0(s)-x_0(s')). \label{eq:S2} 
\end{eqnarray} 
Thus, the native state $x_0(s)$ enters the action in the second term 
of Eq.~(\ref{eq:S2}).  Note that there is no term 
$\beta_0^2V(x_0(s)-x_0(s'))^2$, since it gets cancelled by a similar 
normalization factor for P(\{B\}) in Eq.~(\ref{eq:PB2}). It is useful 
to rewrite Eq.~(\ref{eq:S2}) as 
\begin{eqnarray} 
 &  S_2 = & -\frac{B^2}{4} \int d^dx\,d^dy\,d1\,d2\, 
             A_{12}^{(V)}(x,y) A_{12}^{(\delta)}(x,y) \nonumber \\ 
 &        & -\frac{\beta_0B^2}{2} \int d^dx\,d^dy\,d1\, 
             A_{10}^{(V)}(x,y) A_{10}^{(\delta)}(x,y) 
\end{eqnarray} 
with the notation $A_{12}^{(f)}(x,y) = \int ds f(\Phi(s,1)-x) 
f(\Phi(s,2)-y)$, $A_{10}^{(f)}(x,y) = \int ds f(\Phi(s,1)-x) 
f(x_0(s)-y)$; $f\in\{V,\delta\}$. In the long-chain limit, as 
discussed in Paper I (and references therein), one obtains a 
self-consistent field theoretic formulation, with $S_2$ simplified to, 
\begin{eqnarray} 
 && S_2[\Phi,x_0] = \frac{B^2}{4} \int d^dx d^dy d1 d2 
    \left[ 
      \langle A_{12}^{(V)}(x,y) \rangle 
      \langle A_{12}^{(\delta)}(x,y) \rangle - 
    \right. 
    \nonumber \\ 
 && \left. 
      - A_{12}^{(V)}(x,y) \langle A_{12}^{(\delta)}(x,y) \rangle 
      - \langle A_{12}^{(V)}(x,y) \rangle  A_{12}^{(\delta)}(x,y) 
    \right] \nonumber \\ 
 && + \frac{\beta_0B^2}{2} \int d^dx d^dy d1 
    \left[ 
      \langle A_{10}^{(V)}(x,y) \rangle 
      \langle A_{10}^{(\delta)}(x,y) \rangle 
    \right. 
    \nonumber \\ 
 && \left. 
      - A_{10}^{(V)}(x,y)  \langle A_{10}^{(\delta)}(x,y) \rangle 
      - \langle A_{10}^{(V)}(x,y) \rangle  A_{10}^{(\delta)}(x,y) 
    \right]. 
  \label{eq:S'} 
\end{eqnarray} 
All averages of the type $\langle A^{(V,\delta)} \rangle$ have to be 
calculated self-consistently with $S[\Phi]=S_1[\Phi]+S_2[\Phi]$. (We 
have abbreviated the double average $\langle \langle.\rangle_T 
\rangle_B$ simply by $\langle.\rangle$.) In the limit 
$N\rightarrow\infty$ Eqs.~(\ref{avO}) and (\ref{eq:S'}) provide an 
exact description of the dynamics for an arbitrary native state 
$x_0(s)$.

\section{Average over native state conformations}

It is impossible to solve the model for a general native state 
configuration $x_0(s)$.  We therefore consider a distribution of 
native states and perform the average 
\begin{equation} 
 \overline{<O[\Phi,x_0]>}=\int Dx_0 <O[\Phi,x_0]> \e^{-S_0[x_0]}, 
\end{equation} 
where $S_0[x_0]$ weights each native state conformation in the 
ensemble as 
\begin{equation} 
  S_0[x_0]=1/2\int ds x_0(s) K^{ss'}_{00} x_0(s'), 
\end{equation} 
with 
\begin{equation} 
  K^{ss'}_{00} \equiv \delta_{ss'} ( \mu_0-\partial^2/\partial s'^2). 
\end{equation} 
The parameter $\mu_0$ fixes a size of the globule in this 
ensemble, 
\begin{equation} 
  \langle x_0(s)^2 \rangle = \frac{1}{2\sqrt{\mu_0}} 
  \label{x0^2} 
\end{equation} 
Since the polymer ends are joined, there is translational 
invariance along the coordinate $s$ and $\langle x_0(s)^2\rangle$ 
does not depend on $s$.  Thus, with this procedure, the dynamical 
generating functional for the problem is calculated as 
\begin{equation} 
  e^{-F_{dyn}}=\int Dx_0 D\Phi e^{ -(S_0[x_0]+S_1[\Phi]+S_2[\Phi,x_0]) 
                 }. 
  \label{Fdyn} 
\end{equation} 
There is some formal similarity between the dynamical functional 
$F_{dyn}$ and the static replica partition function.  The integration 
over $Dx_0$ enters in the same way as the extra replica in the static 
formalism.

\section{Correlation functions} 
 
The SUSY correlation functions 
\begin{eqnarray} 
  & & G_{12}^{ss'} \equiv \langle \Phi(s,1) \Phi(s',2) \rangle \label{G12} \\ 
  & & G_{10}^{ss'} \equiv \langle \Phi(s,1) x_0(s') \rangle \label{G10} \\ 
  & & G_{00}^{ss'} \equiv \langle x_0(s)x_0(s') \rangle \label{G00} 
\end{eqnarray} 
contain all the information we are interested in. 
 
$G_{12}^{ss'}$ encodes 16 correlation functions, out of which only 
two, correlation and response function, are independent and nonzero: 
\begin{eqnarray} 
  & G_{12}^{ss'} = & C(s,t_1;s',t_2) 
    + (\bar\theta_2-\bar\theta_1) \times \nonumber \\ 
  & & \times  [ \theta_2 R(s,t_1;s',t_2) - \theta_1 R(s',t_2;s,t_1) ], 
\end{eqnarray} 
with 
\begin{eqnarray} 
  & & C(s,t;s',t') \equiv \langle x(s,t)x(s',t') \rangle, \\ 
  & & R(s,t;s',t') \equiv \langle x(s,t) \tilde x(s',t') \rangle 
        = \frac{\delta\langle x(s,t)\rangle}{\delta h(s',t')}. 
\end{eqnarray} 
The field $h(s',t')$ entering the description of response function is 
an arbitrary external field that couples to $x(s',t')$. The fact that 
only two correlation functions survive is related to Ward identities 
originating from SUSY invariance of the original action $S$. 
 
The supersymmetry of the theory is associated with equilibrium. 
One of the Ward identities resulting from SUSY is the 
fluctuation-dissipation theorem (FDT) which relates correlation 
and response functions.  In the present case, the glassy state 
manifests itself as a spontaneous breaking of supersymmetry, 
leading to a modified FDT, as in previous treatments of other 
models \cite{Kur,CKD}. 
 
$G_{10}^{ss'}$ describes the overlap with the native state.  Due 
to Ward identities, only a single correlation function survives 
(see Appendix A for details): 
\begin{equation} 
  G_{10}^{ss'} = \langle x(s,t)x_0(s') \rangle \equiv \phi(s,t_1;s'). 
  \label{G10a} 
\end{equation} 
Similarly, the native state ensemble is described by 
\begin{equation} 
  G_{00}^{ss'} = \langle x_0(s)x_0(s') \rangle \equiv \Gamma(s;s'). 
\end{equation} 
$G_{12}^{ss'}$ alone is sufficient to describe the RHP model. 
Here we need the two extra functions $G_{10}^{ss'}$ and $G_{00}^{ss'}$. 
 
Also, in what follows, we exploit the translational invariance 
along the $s$ coordinate and define Fourier transforms of all 
correlation functions: $X(s,s')= \int \frac{dk}{2\pi} e^{ik(s-s')} 
X_k$ where $X=C,R,\phi,\Gamma$.

\section{Equations of Motion}

To solve the model we proceed by making a Gaussian variational ansatz 
(GVA), assuming that the fields $\Phi$ are described by the 
approximate action 
\begin{eqnarray} 
 & S_{var} = & 
    \frac{1}{2} \int d1 ds d2 ds' 
    \Phi(s,1) (G^{-1})_{12}^{ss'} \Phi(s',2) + \nonumber \\ 
 && + \int d1 ds ds' 
    \Phi(s,1) (G^{-1})_{10}^{ss'} x_0(s') + \nonumber \\ 
 && + \frac{1}{2} \int ds ds' 
    x_0(s) (G^{-1})_{00}^{ss'}  x_0(s'). 
  \label{Svar} 
\end{eqnarray} 
Technically, this implies the following approximation for $F_{dyn}$: 
\begin{equation} 
  F_{dyn}\approx \langle S \rangle_{var} + F_{var}. 
  \label{Fdyn1} 
\end{equation} 
where 
\begin{eqnarray} 
  & & \e^{-F_{var}}\equiv \int Dx_0 D\Phi \e^{-S_{var}} =\e^{(d/2)Tr\ln G}, \\ 
  & & \langle . \rangle_{var}=\e^{F_{var}}\int Dx_0 D\Phi( . )\e^{-S_{var}}. 
\end{eqnarray} 
The stationarity condition 
\begin{equation} 
  \frac{\delta F_{dyn}}{\delta G_{12}^{ss'}} = 0 
  \label{dFdG} 
\end{equation} 
translates into the equation of motion for Green's function 
$G_{12}^{ss'}$ (see Eqs.~\ref{G12k}-\ref{G00k}). We have derived 
identical equations of motion by using the approach of Ref. \cite{CD}, 
where standard field theoretic identities (e.g. $\langle \Phi \delta 
S/\delta\Phi \rangle=0$ ) are used.  It can be shown that for 
quadratic $S_{var}$ the two procedures give the same result. We omit 
this analysis here to save space. 
 
In a corresponding equilibrium problem, the stationarity condition 
is also an extremum condition and provides a bound on the free 
energy.  Here, since $F_{dyn}$ contains integrations over complex 
fields and Grassmann variables, the GVA does not give a bound on 
$F_{dyn}$.  Nevertheless, it is the first step in a systematic 
approximation procedure, as outlined in Appendix B. 
 
The GVA has been  applied to the problem of a manifold in a random 
potential, in both statics \cite{MP1,MP2} and dynamics 
\cite{CKD,CD}. The method is exact when the dimensionality of the 
manifold is infinite but is only approximate for finite 
dimensionality.  Nevertheless, even for rather low dimensionality 
it has been shown to be a very good approximation in the 
random-manifold problem, where it has been checked numerically 
\cite{CD}.  We have shown in Paper I that the present model is 
closely related to the random-manifold problem. Thus, we hope that 
the GVA  will also be reasonable here, although we have not 
strictly checked its validity. 
 
Using (\ref{Fdyn1}), (\ref{Svar}) and (\ref{eq:SSUSY}) gives the following 
expression for $F_{dyn}$: 
\begin{eqnarray} 
   & & F_{dyn} = \frac{d}{2} \int dsds' K^{ss'}_{00} G^{ss'}_{00} 
             + \nonumber \\ 
   & & + \frac{d}{2} \int dsds'd1d2K^{ss'}_{12}G^{ss'}_{12} 
       - \frac{d}{2} Tr\ln G \nonumber \\ 
   & & - \frac{B^2}{4} 
        \int d^dxd^dy d1d2 \langle A_{12}^{(V)}(x,y) \rangle 
                           \langle A_{12}^{(\delta)}(x,y) \rangle 
       \nonumber \\ 
   & &  - \frac{\beta_0 B^2}{2} 
        \int d^dxd^dy d1d2 \langle A_{10}^{(V)}(x,y) \rangle 
                           \langle A_{10}^{(\delta)}(x,y) \rangle , 
   \label{Fdyn2} 
\end{eqnarray} 
where all averages are to be calculated using $S_{var}$ (see 
Eq.~\ref{Svar}). Performing averages, the fourth and fifth term on the 
right hand side of (\ref{Fdyn2}) become 
\begin{eqnarray} 
  & & F_{dyn}^{(4)} = 
        - \frac{d}{2N} 
            \int d1d2dsds'{\cal V}\left[ (B_{12}^s+B_{12}^{s'})/2 \right], \\ 
  & & F_{dyn}^{(5)} = 
        - \frac{\beta_0 d}{N} 
            \int d1d2dsds'{\cal V}\left[ (B_{10}^s+B_{10}^{s'})/2 \right], 
\end{eqnarray} 
where 
\begin{eqnarray} 
  & & B_{12}^s = \langle [ \Phi(s,1)-\Phi(s,2) ]^2  \rangle = 
                 G_{11}^{ss}  + G_{22}^{ss} - 2 G_{12}^{ss}, \\ 
  & & B_{10}^s = \langle [ \Phi(s,1)-x_0(s) ]^2  \rangle = 
                 G_{11}^{ss}  + G_{00}^{ss} - 2 G_{10}^{ss}, 
\end{eqnarray} 
and 
\begin{equation} 
  {\cal V}(z) = - \frac{\tilde B^2}{d}(z+\sigma)^{-d/2}\ , \ \ 
  \tilde B^2 = \frac{B^2}{2} \frac{N}{v} (4\pi)^{-d/2}. 
\end{equation} 

Performing the variational ansatz (i.e., evaluating Eq.~\ref{dFdG}) 
results in the following equations of motion: 
\begin{eqnarray} 
 & &  \left[ T(\mu+k^2)-D^{(2)}_1 \right] G_{12}^k = \delta_{12} 
       +  2 \int d3 {\cal V}'(B_{13}) \times \nonumber \\ 
 & & \times (G_{32}^k-G_{12}^k) 
     + 2 \beta_0 {\cal V}'(B_{10}) (G_{02}^k-G_{12}^k), 
 \label{G12k} 
\end{eqnarray} 
\begin{eqnarray} 
  &  & \left[ T(\mu+k^2)-D^{(2)}_1 \right] G_{10}^k = 
           2 \int d2 {\cal V}'(B_{12}) \times \nonumber \\ 
  & & (G_{20}^k-G_{10}^k) + 2 \beta_0 {\cal V}'(B_{10})(G_{00}^k-G_{10}^k), 
\end{eqnarray} 
\begin{eqnarray} 
  & ( \mu_0 + k^2) G_{01}^k = & 2 \beta_0 \int d2 {\cal V}'(B_{20}) 
      (G_{21}^k-G_{01}^k), 
\end{eqnarray} 
\begin{equation} 
  ( \mu_0 + k^2) G_{00}^k = 1 + 2 \beta_0 \int d1 {\cal V}'(B_{10}) 
      (G_{10}^k-G_{00}^k), 
  \label{G00k} 
\end{equation} 
and after disentangling the SUSY notation one gets (see Paper I for 
related details) 
\begin{eqnarray} 
 & [T & (\mu+k^2)+\partial/\partial t] C_k(t,t') =  \nonumber \\ 
 & &  2 T R_k(t',t) 
         + 2 \int_{0}^{t} dt'' {\cal V}'\left[B(t,t'')\right] R_k(t',t'') 
      \nonumber \\ 
 & &  + 4 \int_{0}^{t} dt'' {\cal V}''\left[B(t,t'')\right] 
           r(t,t'') \left[ C_k(t,t')-C_k(t'',t') \right] \nonumber \\ 
 & &  - 2 \beta_0 {\cal V}'[A(t)] [ C_k(t,t')-\phi_k(t') ] , 
 \label{Ck} 
\end{eqnarray} 
\begin{eqnarray} 
  & [T & (\mu+k^2)+\partial/\partial t] R_k(t,t') = \delta(t-t') + \nonumber \\ 
  & & + 4 \int_{0}^{t} dt'' {\cal V}''\left[B(t,t'')\right] r(t,t'') \left[ 
    R_k(t,t')-R_k(t'',t') \right] \nonumber \\ 
  & & - 2 \beta_0 {\cal V}'[A(t)]R_k(t,t'), 
  \label{Rk} 
\end{eqnarray} 
\begin{eqnarray} 
 & [ T & (\mu+ k^2)+\partial/\partial t] \phi_k(t) =  \nonumber \\ 
 & &    4 \int_{0}^{t} dt'' {\cal V}''\left[B(t,t'')\right] 
          r(t,t'') \left[ \phi_k(t)-\phi_k(t'') \right] \nonumber \\ 
 & &    + 2 \beta_0 {\cal V}'[A(t)] (\Gamma_k-\phi_k(t)), 
 \label{Phik1} 
\end{eqnarray} 
\begin{eqnarray} 
 & & (\mu_0+k^2) \phi_k(t) = 
       2 \beta_0 \int_{0}^{t}dt''{\cal V}'[A(t'')] R_k(t,t''), 
 \label{Phik2} 
\end{eqnarray} 
\begin{eqnarray} 
 & & (\mu_0+k^2) \Gamma_k = 1, 
\end{eqnarray} 
where $B(t,t')$ and $A(t)$ are defined as $B(t,t') = \langle (x(s,t) - 
x(s,t'))^2 \rangle = C(s,t;s,t) + C(s,t';s,t') - 2 C(s,t;s,t')$ and 
$A(t) = \langle (x(s,t) - x_0(s))^2 \rangle = C(s,t;s,t) - 
2\phi(s,t;s) + \Gamma(s,s)$.  Note that due to translational 
invariance with respect to $s$ both $B(t,t')$ and $A(t)$ are 
$s$-independent.  The equations of motion for $C_k(t,t')$ and 
$R_k(t,t')$ are almost identical to those for the pure RHP 
model. Coupling to the native state enters through the terms 
proportional to $\beta_0$. Again, for large selection temperature 
$\beta_0\rightarrow 0$ and one recovers the RHP model.

\section{Extracting order parameters} \label{sec:order_parameters}

The equations of motion are coupled integro-differential equations 
with initial conditions given by $C_k(0,0)$, $\phi(0)$, and (we use 
Ito's convention) $R(t+\epsilon,t)\rightarrow 1$ as 
$\epsilon\rightarrow 0$.  To solve the equations analytically we have 
to consider several assumptions (which can be checked by numerical 
solution). 
 
First, we make the (rather strong) standard assumptions from aging 
theory for spin glasses about the asymptotic behavior of the 
solutions: In the regime of time translational invariance (TTI), 
\begin{eqnarray} 
  & &  \lim_{t\rightarrow\infty} C_k(t+\tau,t) = C_k(\tau), \\ 
  & &  \lim_{t\rightarrow\infty} R_k(t+\tau,t) = R_k(\tau), 
\end{eqnarray} 
and, in the aging regime, 
\begin{eqnarray} 
  & & \lim_{t\rightarrow\infty} C_k(t,\lambda t) = 
        q_k \hat C_k(\lambda), \\ 
  & & \lim_{t\rightarrow\infty} R_k(t,\lambda t) = 
        \frac{1}{t} \hat R_k(\lambda). 
\end{eqnarray} 
The validity of these assumptions could be checked numerically. Since 
this has been done for equations of similar type elsewhere~\cite{CD}, 
we omit it in the present analysis. 
 
Second, it is well known that asymptotic solutions of such equations 
can be characterized by a few order parameters 
\cite{CKD,CD,CK1,BCKP,CK2}. They are defined as 
\begin{eqnarray} 
 & & \tilde q_k = \lim_{t\rightarrow\infty} C_k(t,t), \\ 
 & & q_k = \lim_{\tau\rightarrow\infty} C_k(\tau), \\ 
 & & q_{0,k} = \lim_{\lambda\rightarrow 0} q_k \hat C_k(\lambda), \\ 
 & & \varphi_k=\lim_{t\rightarrow\infty}\phi_k(t). 
\end{eqnarray} 
The following $k$-integrated quantities will also be useful: 
\begin{eqnarray} 
 & & \tilde q \equiv \int \frac{dk}{2\pi} \tilde q_k = 
       \lim_{t\rightarrow\infty} \langle x(s,t)x(s,t) \rangle , \\ 
 & & q \equiv \int \frac{dk}{2\pi} q_k = 
       \lim_{\tau\rightarrow\infty} 
       \lim_{t\rightarrow\infty} \langle x(s,t)x(s,t+\tau) \rangle ,\\ 
 & & q_0 \equiv \int \frac{dk}{2\pi} q_{0,k} = 
       \lim_{\lambda\rightarrow 0} 
       \lim_{t\rightarrow\infty} \langle x(s,t)x(s,\lambda t) \rangle ,\\ 
 & & \varphi \equiv \int \frac{dk}{2\pi} \varphi_k = 
       \lim_{t\rightarrow\infty} \langle x(s,t) x_0(s) \rangle. 
\end{eqnarray} 
$\tilde q$ measures the size of the globule, $q$ measures the 
persistent correlation in the TTI regime, $q_0$ the asymptotic 
correlation in the aging regime, and $\varphi$ the overlap with native 
state. Also, it is useful to define 
\begin{eqnarray} 
  & & b = 2 ( \tilde q - q ) \, , \,\,\, b_0 = 2 ( \tilde q - q_0 ), 
          \label{bqq0} \\ 
  & & a \equiv \lim_{t\rightarrow\infty} 
               \langle [x(s,t)-x_0(s)]^2 \rangle 
               = \tilde q - 2 \varphi + \frac{1}{2\sqrt{\mu_0}}. 
\end{eqnarray} 

Third, we assume that the generalized fluctuation dissipation theorem 
is valid in the form 
\begin{equation} 
  \hat R_k(\lambda) = \frac{x}{T} 
                        \, q_k \frac{ d\hat C_k(\lambda) }{ d\lambda }, 
  \label{GFDT} 
\end{equation} 
$x$ could in principle depend on $k$ and $C_k$. However, related 
models have been studied in detail and they exhibit one step replica 
symmetry breaking with a $k$-independent $x$.  This one step replica 
symmetry breaking ansatz in our dynamical study translates exactly to 
Eq.~(\ref{GFDT}).

\section{Relating order parameters} 
 
For $t=t'$ and $t\rightarrow\infty$ Eq.~(\ref{Ck}) gives 
\begin{eqnarray} 
   & & T(\mu+k^2) \tilde q_k =  T 
      + \frac{2}{T} {\cal V}'(b) (1-x) ( \tilde q_k - q_k ) \nonumber \\ 
   & & \ \ \ \ \ \ 
        + \frac{2}{T} {\cal V}'(b_0) x (  \tilde q_k - q_{0,k}) 
        -2\beta_0 {\cal V}'(a)(\tilde q_k - \phi_k). 
   \label{eq:qtk2} 
\end{eqnarray} 
With $t=t'+\tau$ and $t'\rightarrow\infty$ and then 
$\tau\rightarrow\infty$ Eq.~(\ref{Ck}) becomes 
\begin{eqnarray} 
   & & T(\mu+k^2) q_k = 
      \frac{2}{T} ( {\cal V}'(b) 
        - x {\cal V}'(b_0) ) ( \tilde q_k - q_k ) \nonumber \\ 
   & &  \ \ \ \ \ \ 
       + \frac{2}{T} {\cal V}'(b_0) x ( \tilde q_k - q_{0,k} ) 
       -2\beta_0 {\cal V}'(a)(q_k - \phi_k). 
   \label{eq:qk2} 
\end{eqnarray} 
Eq.~(\ref{Ck}) in the aging regime $t'=\lambda t$, first for 
$t\rightarrow\infty$ and then $\lambda\rightarrow 0$, gives 
\begin{eqnarray} 
   & & T(\mu+k^2) q_{0,k} = 
      \frac{2}{T} {\cal V}'(b_0) (1-x) ( \tilde q_k - q_k )  \nonumber \\ 
   & & \ \ \ \ \ \ 
       + \frac{2}{T} {\cal V}'(b_0) x ( \tilde q_k - q_{0,k} ) 
       -2\beta_0 {\cal V}'(a)(q_{0,k} - \phi_k). 
   \label{eq:q0k2} 
\end{eqnarray} 
Eqs.~(\ref{Phik1}) and (\ref{Phik2}) result in two equations for 
$\varphi_k$, 
\begin{eqnarray} 
  & T ( \mu + k^2 ) \varphi_k = & 
      \frac{2}{T_0} {\cal V}'(a) ( \Gamma_k - \varphi_k ) 
  \label{varphik1} \\ 
  & (\mu_0 + k^2 ) \varphi_k = & 
      \frac{2}{TT_0} {\cal V}'(a) x ( q_k - q_{o,k} ) \nonumber \\ 
  & &   + \frac{2}{TT_0} {\cal V}'(a) ( \tilde q_k - q_k ) 
  \label{varphik2} 
\end{eqnarray} 
They are equivalent; one can chose to solve for the order parameters 
working with either (\ref{varphik1}) or (\ref{varphik2}). This seems a 
rather remarkable coincidence. We believe that it originates from the 
SUSY invariance of the original action $S$. For example, a similar 
comment holds for equations (\ref{Ck}) and (\ref{Rk}); they are 
equivalent in the TTI regime and one can derive one from the other. 
The `conspiracy' of (\ref{Phik1}) and (\ref{Phik2}) not contradicting 
each other is very likely a similar phenomenon. Eq.(\ref{Rk}) for 
$\lambda=1$ reduces to 
\begin{equation} 
  \hat R_k(1) (\tilde\mu+k^2+\Sigma)  = - (\tilde q_k - q_k ) 
  \frac{4{\cal V}''(b)}{T^2} \hat r(1), 
  \label{MSCk} 
\end{equation} 
where 
\begin{equation} 
\hat r(\lambda) \equiv \int \frac{dk}{2\pi} \hat R_k(\lambda) 
\end{equation} 
and $\Sigma$ is defined by 
\begin{equation} 
  \Sigma = x \frac{2}{T^2} \left( {\cal V}'(b) - {\cal V}'(b_0) \right). 
  \label{Sigma} 
\end{equation} 

Solving these equations for the order parameters gives 
\begin{eqnarray} 
  & & b = \frac{1}{\sqrt{\tilde\mu+\Sigma}}, \label{qtq} 
\end{eqnarray} 
\begin{eqnarray} 
  & & b_0 = 
      \frac{1}{x} \frac{1}{\sqrt{\tilde\mu}} 
      + \frac{x-1}{x} \frac{1}{\sqrt{\tilde\mu+\Sigma}} 
  \label{qq0}, 
\end{eqnarray} 
\begin{eqnarray} 
  & \tilde q = & \frac{b_0}{2} 
         + \frac{{\cal V}'(b_0)}{4T^2\tilde\mu^{3/2}} 
         + \frac{1}{4\sqrt{\mu_0}} 
           \left( 
              \frac{1-\frac{\mu}{\tilde\mu}}{1-\frac{\mu_0}{\tilde\mu}} 
           \right)^2 \times \nonumber \\ 
  & &       \times \left(2+\sqrt{\frac{\mu_0}{\tilde\mu}} \right) 
           \left( 1 -\sqrt{\frac{\mu_0}{\tilde\mu}} \right)^2, 
  \label{qt} 
\end{eqnarray} 
\begin{eqnarray} 
  & a = & \frac{b_0}{2} 
         + \frac{{\cal V}'(b_0)}{4T^2\tilde\mu^{3/2}} 
         + \frac{1}{4\sqrt{\mu_0}} 
           \frac{1}{ 
             \left( 1+\sqrt{\frac{\mu_0}{\tilde\mu}} \right)^2 
           } \times \nonumber \\ 
  & &       \times 
           \left[ 
              \sqrt{\frac{\mu_0}{\tilde\mu}} 
              \left( 1+2\sqrt{\frac{\mu_0}{\tilde\mu}} \right) 
              + 2 \frac{\mu}{\tilde\mu}\sqrt{\frac{\mu_0}{\tilde\mu}} 
           \right. 
           \nonumber \\ 
  & &      \left. 
           \,\,\,\,\,\,\,\,\,\,  + \left( \frac{\mu}{\tilde\mu} \right)^2 
             \left( 2+\sqrt{\frac{\mu_0}{\tilde\mu}} \right) 
           \right], 
  \label{a} 
\end{eqnarray} 
\begin{eqnarray} 
  & & \tilde\mu = \mu + \frac{2}{TT_0}{\cal V}'(a), \label{mut} 
\end{eqnarray} 
and the combination of Eq.~(\ref{qtq}) and (\ref{MSCk}) gives 
\begin{eqnarray} 
  && 0 = \hat r(1) \left[ T^2 + b^3 {\cal V}''(b) \right].  \label{MSC} 
\end{eqnarray} 
Furthermore, the overlap $\varphi$ with the native state 
is given by 
\begin{equation} 
  \varphi = \frac{1}{2\sqrt{\mu_0}} 
    \frac{ 
       1-\frac{\mu}{\tilde\mu} 
         }{ 
       1+\sqrt{ 
           \frac{\mu_0}{\tilde\mu} 
         } 
    }. 
  \label{varphi} 
\end{equation} 
All overlap order parameters are positive.  However, this result is 
not obvious and has to be obtained after some algebra. 
 
These equations have two kinds of solutions.  In one kind, $b=b_0$, so 
there is no glassiness (aging).  For this kind of solution, the 
parameter $x$ is irrelevant.  We call such solutions ``ergodic''. 
(While it will turn out that some of them are not truly ergodic, in 
the sense of describing states where the entire configuration space is 
visited with Boltzmann probabilities, they violate ergodicity in a 
rather trivial way, like a ferromagnet below the Curie temperature. 
We could call them ``non-glassy'', but we prefer not to use a negative 
term.) 
 
For an ergodic solution, with $b=b_0$, $\Sigma=0$.  Furthermore, $\hat 
r(\lambda)=0$, so Eqn.~(\ref{MSC}) is trivially satisfied.  One then 
has to solve the four equations (\ref{qtq}) and (\ref{qt}-\ref{mut}) 
for $b$, $\tilde q$, $a$ and $\tilde \mu$. 
 
The stability of such a phase against glassiness can be determined 
using the analysis we presented in Paper I (see Fig.~1). There, we 
studied a model with no native-state bias in its interactions ($T_0 = 
\infty$) for finite $\mu$.  The boundary of the glassy state as a 
function of $\mu$ has a form qualitatively like that in the p-spin 
glass as a function of field \cite{CS,CHS}.  In the present model, the 
presence of the native state enters the calculation solely through the 
replacement of $\mu$ by $\tmu$.  Therefore, if a particular $T$ and 
$\tmu$ fall in the glassy regime (the region below the full and dashed 
lines) in Fig.~1, the ergodic ansatz has to be given up. 
 
The instability can occur in two ways, according to whether $\tmu$ is 
bigger or smaller than the critical value $\tmu_c$.  Above $\tmu_c$, 
the line separating glassy from ergodic regions is an Almeida-Thouless (AT) 
line; below it the stability condition 
\begin{equation} 
T^2 + b^3 {\cal V}''(b) >0, \label{eq:ATinequality} 
\end{equation} 
is violated.  For $\tmu < \tmu_c$, there is no AT instability. The 
transition is like that for the completely random heteropolymer. 
To find such a transition, we have to solve for a glassy phase, 
characterized in part by a value of the FDT-violation parameter $x < 
1$ and then find where in the parameter space $x \rightarrow 1$.  In 
the region where the $x<1$ solution exists, the associated ergodic 
phase is unstable and is replaced by the glassy one. 
 
In a glassy phase, aging is present: $\hat r(1)\ne 0$, so the quantity 
in brackets in Eq.~(\ref{MSC}) has to vanish, i.e., the AT condition 
has to be satisfied as an equality, rather than an inequality.  This 
so-called marginal stability condition determines $b$ as a function of 
temperature.  In this case we have three more unknowns, $\Sigma$, 
$b_0$ and $x$, making a total of seven, and seven equations, 
(\ref{qtq}-\ref{MSC}), to solve for them. 
 
We look for ergodic solutions first in the next section, and we 
examine their stability.  Then, in the following section, we study 
glassy solutions (within the 1-step aging ansatz of section VII) and 
identify the regions in the parameter space where they hold.

\section{Ergodic phases} \label{sec:erg_phase} 
 
For ergodic phases, Eqns.~(\ref{qtq}-\ref{mut}) reduce to 
\begin{eqnarray} 
  && b=b_0=\frac{1}{\sqrt{\tilde\mu}} \label{berg} \\ 
  & \tilde q = & \frac{1}{2\sqrt{\tilde\mu}} 
         + \frac{{\cal V}'(1/\sqrt{\tilde\mu})}{4T^2\tilde\mu^{3/2}} 
         + \frac{1}{4\sqrt{\mu_0}} 
           \left( 
              \frac{1-\frac{\mu}{\tilde\mu}}{1-\frac{\mu_0}{\tilde\mu}} 
           \right)^2 \times \nonumber \\ 
  & &       \times \left(2+\sqrt{\frac{\mu_0}{\tilde\mu}} \right) 
           \left( 1- \sqrt{\frac{\mu_0}{\tilde\mu}} \right)^2 
           \label{qterg} \\ 
  & a = & \frac{1}{2\sqrt{\tilde\mu}} 
         + \frac{{\cal V}'(\frac{1}{\sqrt{\tilde\mu}})}{4T^2\tilde\mu^{3/2}} 
         + \frac{1}{4\sqrt{\mu_0}} 
           \frac{1}{ 
             \left( 1+\sqrt{\frac{\mu_0}{\tilde\mu}} \right)^2 
           } \times \nonumber \\ 
  & &       \times 
           \left[ 
              \sqrt{\frac{\mu_0}{\tilde\mu}} 
              \left( 1+2\sqrt{\frac{\mu_0}{\tilde\mu}} \right) 
              + 2 \frac{\mu}{\tilde\mu}\sqrt{\frac{\mu_0}{\tilde\mu}} 
           \right. 
           \nonumber \\ 
  & &      \left. 
           \,\,\,\,\,\,\,\,\,\,  + \left( \frac{\mu}{\tilde\mu} \right)^2 
             \left( 2+\sqrt{\frac{\mu_0}{\tilde\mu}} \right) 
           \right] 
           \label{aerg} \\ 
  & & \tilde\mu = \mu + \frac{2}{TT_0} {\cal V}'(a) \label{muterg} 
\end{eqnarray} 
They can be solved numerically: given $\mu$, $\mu_0$, $T$ and $T_0$ 
one can find $\tilde\mu$, which in turn determines $\tilde q$, $b=b_0$ 
(equivalently $q=q_0$), and $\varphi$. However, it is possible to gain 
some analytic understanding in a few soluble limits. 
 
In this discussion we will concentrate on the limit of small $\mu$. 
As we noted in paper I, if we want to confine $N$ monomers within a 
gyration radius $\sqrt{\tilde q} \propto \mu^{-1/4}$, we need $\mu 
\propto N^{-4/d}$.  Thus, for a long polymer $\mu \rightarrow 0$.  We 
will also take $\mu=\mu_0$ to simplify the algebra a bit. 
 
The pair of equations (\ref{aerg}) and (\ref{muterg}) fully determine 
$\tmu$ as function of $T$ and $T_0$.  For $\mu_0=\mu$ they take the 
form 
\bea 
a(\tmu) &=& \frac{1}{2\sqrt{\tmu}} +\frac{\Bt^2}{8T^2 \tmu^{3/2} 
(\sigma + \tmu^{-1/2})^{\frac{d}{2}+1}} \nonumber \\ 
&+& \frac{1}{4\sqrt{\tmu}}\left( 1 + \frac{\mu}{\tmu}\right) 
\lb{eq:a}   \\ 
\tmu(a) &=& \mu + \frac{\Bt^2}{TT_0 (\sigma + a)^{\frac{d}{2}+1}} 
\lb{eq:tmu} 
\eea 
Given $\tmu$, $T$, $T_0$ one can find the overlap with the native 
state $\varphi$ and the size of the polymer from (\ref{varphi}).

\subsection{Random-globule state} 
 
It is immediately evident that when both the temperature $T$ and the 
selection temperature $T_0$ are large, $\tilde \mu \approx \mu$ in 
(\ref{muterg}), leading to a random-globule solution $a = b = 
\mu^{-1/2}$, $\tilde q = \mu^{-1/2}/2$, $\varphi = 0$.  What is not so 
obvious is that in the $\mu \rightarrow 0$ limit a solution very close 
to this exists all the way down to very low temperatures, even for 
small $T_0$.  In this subsection we examine this state in detail. 
 
We look first for solutions of Eqs.~(\ref{eq:a}) and (\ref{eq:tmu}) 
with the ansatz $\alpha \equiv \tmu/\mu$ fixed and $\mu\rightarrow 
0$. We call this the random globule ansatz, since, as will be shown, 
the polymer does not have any fixed conformation (it is melted), and 
on the average the conformations it adopts have zero overlap with the 
native state.  (Strictly speaking, this is the only truly ergodic 
phase we find.)  For $a$ we get, 
\begin{equation} 
  a = \frac{1}{\sqrt{\mu}} 
      \left[ 
        \frac{ 
              3+\frac{1}{\alpha} 
         }{ 
              4\sqrt{\alpha} 
         } 
+ 
        {\cal O}(\mu^{(d-2)/2}) 
      \right], 
\end{equation} 
which, after inserting into (\ref{eq:tmu}), gives 
\begin{equation} 
  \alpha \approx 1 + \frac{\Bt^2}{TT_0} \mu^{(d-2)/4} 
                 \left( 
                   \frac{ 
                         4 \sqrt{\alpha} 
                     }{ 
                         3 + \frac{1}{\alpha} 
                   } 
                 \right)^{d/2+1}. 
  \label{alpha} 
\end{equation} 
Eqn.~(\ref{alpha}) can be used to calculate $\alpha$ as a function 
of $\mu$. One can see easily that $\alpha\rightarrow 1$ when 
$\mu\rightarrow 0$. This shows that our ansatz is self-consistent 
in the limit of small $\mu$. Also, (\ref{qterg}) and 
(\ref{varphi}) become 
\begin{eqnarray} 
  & & \varphi=\frac{1}{2\sqrt{\mu}} 
              \left[\frac{\alpha-1}{2}+{\cal O}(\alpha-1)^2\right] 
      \label{varphi2} \\ 
  & & \tilde q=\frac{1}{2\sqrt{\mu}}\left[1+{\cal O}(\alpha-1)\right]. 
      \label{qterg2} 
\end{eqnarray} 
 
The normalized overlap between the polymer conformation and the native 
state is: 
\begin{equation} 
   \cos \theta = 
   \frac{ 
     \lim_{t\rightarrow\infty}\langle x(s,t)x_0(s)\rangle 
   }{ 
     \sqrt{ 
       \lim_{t\rightarrow\infty} \langle x(s,t)^2 \rangle 
                              \langle x_0(s)^2 \rangle 
     } 
   } = 
   \frac{\varphi 
           }{ 
     \sqrt{\tilde q (\frac{1}{2\sqrt{\mu}}}) 
    }. 
  \label{costh} 
\end{equation} 
>From (\ref{varphi2}) and (\ref{qterg2}) we get 
$\cos(\theta)\sim(\alpha-1)\sim\mu^{(d-2)/4}$. Thus there is no 
overlap with native state as $\mu\rightarrow 0$. 
 
Furthermore, to check that polymer does not freeze into some other 
conformation, we calculate the normalized overlap between two 
configurations taken at very different times, 
\begin{equation} 
  \cos \theta' = 
   \frac{ 
     \lim_{\tau,t\rightarrow\infty} \langle x(s,t)x(s,t+\tau)\rangle 
    }{ 
      \sqrt{ [\lim_{t\rightarrow\infty} \langle x(s,t)^2 \rangle ]^2 } 
    } = 
   \frac{q}{\tilde q}. 
   \label{costh'} 
\end{equation} 
After rewriting 
\begin{equation} 
  q/\tilde q=1-\frac{b}{2\tilde q}= 
  1-\frac{1}{2\sqrt{\tmu\tilde q}}, 
  \label{q/qt} 
\end{equation} 
and, using (\ref{qterg2}), we get $\cos \theta'={\cal 
O}(\alpha-1)$. Again, as $\mu\rightarrow 0$, $\cos \theta' \rightarrow 
0$. This confirms that the ansatz $\alpha ={\cal O}(1)$ and $\mu 
\rightarrow 0$ leads to a melted random-globule-like phase.  This 
phase is identical to that found at high temperatures for the 
completely random heteropolymer in paper I. 
 
The validity of the present ansatz rests upon the fact that we can 
solve Eq.~(\ref{alpha}). Clearly, for $\mu\rightarrow 0$ a solution 
can always be found, namely $\alpha=1$.  Since the physically relevant 
$\mu$ is $\propto N^{-4/d}$, we can always satisfy this equation, for 
any $T_0$, in the limit $N \rightarrow\infty$. 
 
We now address briefly the question of what happens for finite $N$ 
(and $\mu$).  One can easily see that Eq.~(\ref{alpha}) has two 
solutions when $\mu^{(d-2)/4}/(TT_0)$ is not too large (e.g. by 
plotting the left- and right-hand side as functions of $\alpha$).  The 
solution close to 1 is lost when the slopes of the left- and 
right-hand sides become roughly equal.  Evaluating these slopes leads 
to the condition 
\be 
\frac{3\Bt^2}{4TT_0}\left(\frac{d}{2}+1\right)\mu^{\frac{d-2}{4}} <1 
\label{eq:coilcond} 
\ee 
for the existence of a random-globule-like state. 
 
Some caution is in order. Working this out for finite $N$, $d=3$, and 
an average density of $1$, we find that the inequality 
(\ref{eq:coilcond}) is violated below a temperature 
\be 
T_x = \left( \frac{\pi}{6}\right)^{1/2}\frac{15\Bt^2}{8T_0} N^{-1/3}. 
\ee 
With the small power of $N^{-1}$, one has to go to quite large $N$ to 
make this temperature very low.  Thus our statement that the 
random-globule-like state exists for all temperatures in the $\mu 
\rightarrow 0$ limit may be of limited relevance for real 
3-dimensional heteropolymers of the length of typical proteins. 
Nevertheless, here we are just considering this simple limit. 
 
We now discuss the stability of this solution. In the large-$N$ limit, 
it is locally stable against spontaneous formation of a native-like 
state at any $T$ and $T_0$.  However, it is unstable against glass 
formation at low temperatures: Since it is identical with the 
random-globule solution of the completely random heteropolymer 
problem, we can take over the result from paper I that it is unstable 
below a temperature $T_g \propto \Bt$, with the constant of 
proportionality of order 1.  This glass temperature is independent of 
$T_0$.  (In Fig.~1 this is the transition at $\tmu \rightarrow 0$.) 
Thus, wherever the system is in a random-globule-like state at $T > 
T_g$, it will no longer equilibrate if the temperature is lowered 
below $T_g$.  Instead, it will become glassy and its dynamics will 
show aging.

\subsection{Ergodic native state} 
 
At low $T_0$ and $T$, one expects that the polymer should be very 
close to its native state, i.e., small $a$.  Therefore we also look 
for such solutions of Eqs.~(\ref{eq:a}) and (\ref{eq:tmu}).  We will 
try to solve equations (\ref{eq:a}) and (\ref{eq:tmu}) in the limit 
where $\mu \rightarrow 0$ and $\tmu$ stays finite. The limit 
$\mu\rightarrow 0$ turns out not to involve any subtleties when $\tmu$ 
is kept constant, so we will just set $\mu = 0$ from the outset. 
Eqs.~(\ref{eq:a}) and (\ref{eq:tmu}) become 
\begin{eqnarray} 
a(\tmu) &=& \frac{3}{4\sqrt{\tmu}} +\frac{\Bt^2}{8T^2 \tmu^{3/2} 
(\sigma + \tmu^{-1/2})^{\frac{d}{2}+1}} \lb{eq:a1} \\ 
\tmu(a) &=& \frac{\Bt^2}{TT_0 (\sigma + a)^{\frac{d}{2}+1}} 
\lb{eq:tmu1} 
\end{eqnarray} 
These equations can be solved for $\tmu$ as function of $T$ and 
$T_0$. However, one has to keep in mind that $\mu\rightarrow 0$ has 
been taken. This implies that (\ref{varphi}) and (\ref{qterg}) become 
\begin{equation} 
  \varphi \approx \frac{1}{2\sqrt{\mu}} , \ \ \ 
  \tilde q \approx \frac{1}{2\sqrt{\mu}} 
  \label{varphiqt}, 
\end{equation} 
and, inserting (\ref{varphiqt}) into (\ref{costh}), the normalized 
overlap between native state and polymer conformations, becomes $\cos 
\theta \approx 1$.  Furthermore, because of its large overlap with the 
native state, the polymer is essentially frozen.  This can be seen by 
calculating the normalized overlap between two polymer conformations 
after a very long time interval, as in previous section.  Inserting 
(\ref{varphiqt}) into (\ref{costh'}) and (\ref{q/qt}) gives $\cos 
\theta' \approx 1-\sqrt{\mu/\tmu}\rightarrow 1$. 
 
There is interesting behavior associated with the limit 
$\mu\rightarrow 0$ for very long polymers. When the polymer gets 
longer and longer ($N\rightarrow\infty$) a finite part of the chain is 
not in the native state conformation, since $a$ stays constant.  The 
rest of the chain is in the native state, which can be seen from the 
fact that overlap with native state approaches 1.  Thus, in the limit 
of a very long polymer, the fraction of chain not in the native 
state conformation becomes negligible: the recipe for biasing the 
coupling constants $B_{ss'}$ described in chapter II works best for 
long polymers. 
 
In the following we will proceed with the solution of equations 
(\ref{eq:a1}) and (\ref{eq:tmu1}).  Before continuing, it will be 
useful to compactify notation a bit. Making the change of variables 
$\hat X = X / \sigma$ for $X=b,\,b_0,\,a,\,q,\,\tilde q$; $\hat Y = Y 
\sigma^2$ for $Y=\mu,\,\tmu$; and $\hat Z=Z\sigma^{(d-2)/4}/\Bt$ for 
$Z=T,\,T_0$, we get equations of the same form, with $X \rightarrow 
\hat X$, $Y \rightarrow \hat Y$ and $Z \rightarrow \hat Z$, but with 
$\sigma =1$ and $\Bt=1$.  Thus, without loss of generality, we can 
choose units with $\sigma=1$ and $\Bt=1$ (and remove the hats).  From 
now on we do this. 
 
The working strategy for solving the equations is as follows. For 
fixed $T$, one can consider $T_0$ as a function of $\tmu$. This can be 
easily done by inserting the expression for $a$ from Eq.~(\ref{eq:a1}) 
into (\ref{eq:tmu1}), thus writing $\beta_0 = 1/T_0$ as 
\begin{equation} 
  \beta_0(\tmu,T) = T\tmu[1+a(\tmu,T)]^{d/2+1} 
  \label{invt0tmu} 
\end{equation} 
The four panels of Fig.~2 shows the shape of $\beta_0(\tmu,T)$ as a 
function of $\tmu$ for four different temperatures. We want ultimately 
to construct a phase diagram in the $(\beta_0,T)$ plane.  Therefore we 
have to specify $T$ (one panel of the figure) and $\beta_0$ and ask 
whether one or more solutions, i.e., particular values of $\tmu$ which 
solve Eq.~(\ref{invt0tmu}), exist.  For example, in panel (a) in 
Fig.~2, a horizontal line at $\beta_0>\beta_0^{\rm min}$ intersects 
$\beta_0(\tmu,T)$ curve at two places, indicating two solutions 
$\tmu=\tmu_{1}, \tmu_{2}$.  To make the figures more readable we have 
shown such a horizontal line, at a particular values of $\beta_0$, 
only in panel (a). If this horizontal line is moved below 
$\beta_0^{\rm min}$, it will never intersect the $\beta_0(\tmu,T)$ 
curve.  Thus, we can see that for every $T$, there is a value 
$\beta_0^{\rm min}(T)$ below which no solutions exist. 
 
We proceed with the analysis of Fig.~2.  For sufficiently high 
temperatures (panels (a)-(c)) there are exactly two solutions for all 
$\beta_0 > \beta_0^{\rm min}$.  Of these, the one with the larger 
value of $\tmu$ is a stable solution (local free energy minimum) 
describing the ergodic native phase.  For example, the solution 
labeled $\tmu_2$ in panel (a) is of this sort.  The one with the 
smaller value of $\tmu$ (e.g., the one labeled by $\tmu_1$ in panel 
(a)) is unstable.  It describes a free energy maximum between the 
minima at the random-globule and ergodic native states.  We will call 
such states ``unstable stationary'' (abbreviated US).  (We have not 
done a static calculation to show this, but the situation here is 
analogous to that in an ordinary ferromagnet below the Curie 
temperature.  There, one has three solutions of the mean field 
equations, one with positive, one with negative, and one with zero 
magnetization.  The middle one, with zero magnetization, is 
unstable). The US state has a lower overlap with the native 
conformation than the ergodic-native solution does, because it has a 
smaller value of $\tmu$.  As $\beta_0$ is increased from below through 
$\beta_0^{\rm min}$, the native-state and US-state solutions appear 
together and separate.  For the temperatures of panels (a)-(c), they 
both exist for all $\beta_0 >\beta_0^{\rm min}$. 
 
Panel (d) (at the lowest of the temperatures) shows a more complex 
behavior where double-minimum structure appears. We have found 
numerically that this happens below $T\approx 0.20$.  Here the 
behavior around $\beta_0^{\rm min}$ is just as in the other cases, but 
we note that at this temperature $\beta_0(\tmu,T)$ has a second local 
minima at a smaller value of $\tmu$.  Thus there is a range 
$\beta_1^{\rm min}<\beta_0<\beta_1^{\rm max}$ for which there are four 
solutions.  The rightmost one is stable and describes the 
ergodic-native phase, as before.  Moving from right to left, the 
solutions alternate between stability and instability.  Thus the 
second solution from the left represents a locally stable 
conformation.  It is also correlated with the native state, since 
$\tmu$ is finite (though we always find $\tmu \ll 1$ in 3 dimensions). 
The remaining two solutions (with $\partial \beta_0/\partial \tmu <0$) 
represent US states (local free energy maxima) between it and the 
random-globule phase in one direction and the ergodic-native phase in 
the other. 
 
Plotting $\beta_0^{\rm min}$ against $T$, we obtain the stability 
boundary indicated by the thick solid curve in Fig.~3.  Within our 
present assumption of ergodicity, everywhere to the right of this line 
the ergodic-native phase is dynamically stable.  One can invert the 
relation $\beta_0^{\rm min}(T)$, obtaining a transition temperature 
$T_n(\beta_0)$, the maximum temperature for which the ergodic native 
phase is dynamically stable.  It is separated from the (also stable) 
random-globule phase by a barrier, the top of which is described by 
the unstable solution. 
 
In Fig.~3 we also indicate the region in the $(\beta_0,T)$ plane where 
the second locally-stable solution is found.  This region has the form 
of a kind of sliver extending out toward large $\beta_0$ at low 
temperatures. 
 
So far we have not examined the stability of these solutions against 
glassiness.  As indicated above, we do this with the help of Fig.~1: 
Stable solutions can not lie in the range $\tmu_{min} < \tmu < 
\tmu_{AT}$.  In Fig.~2, these limits are marked on the $\tmu$ axes. 
We thus see, for example, in Panel (c), that the native-state 
solutions found for the range of $\beta_0$ corresponding to values of 
$\tmu$ between $\tmu_*$ and $\tmu_{AT}$ are not acceptable: they 
violate the AT stability condition (\ref{eq:ATinequality}). 
 
Similarly, in panel (b) the US solutions found for a range of 
$\beta_0$ values can also be seen to lie in the forbidden region.  And 
the intermediate locally-stable states that we identified in panel (d) 
always lie in a glassy region. 
 
In Fig.~3 we also plot the AT line (\ref{MSC}) in the $(\beta_0,T)$ 
plane, indicating the regions where the various kinds of ergodic 
solutions are forbidden.  For the native-phase solutions, the 
forbidden region is a strip mostly at low values of $T$ (diagonally 
cross-hatched region between thick and AT line).  However, it ``wraps 
around'' at the leftmost part of the region where those solutions are 
found. 
 
The forbidden region for the US solutions occupies most of the region 
where these solutions occur below $T_{max}$, the maximum temperature 
for a glass transition shown in Fig.~1, including the entire portion 
of it below $T_g$, the glass instability temperature of the 
random-globule state. 
 
The structure in a tiny region near the minimum value of $\beta_0$ for 
which ergodic-native solution are found is a bit complicated and 
cannot be seen in the top panel of Fig.~3.  Therefore, the lower panel 
shows an enlargement of this region. 
 
In summary, we have found four kinds of ergodic solutions. One 
essentially describes a random globule state.  It is locally stable 
(in the limit of a large globule) at all $T$ and $\beta_0$ against 
condensation into a native-like state, but unstable against glass 
formation everywhere below a transition temperature $T_g$.  The second 
kind of solution describes a phase which is highly correlated with the 
native state conformation, and it is stable in most of the region 
where the solution exists.  The third kind of solution describes a 
locally stable state, correlated with the native state but more weakly 
so than the ergodic native phase just described.  It is never stable 
against glass formation.  Finally, there are unstable solutions, found 
whenever the ergodic-native solutions exist.  They describe US states, 
free energy maxima between pairs of the previously-described 
solutions.  However, in a large part of the region where these 
solutions are found (roughly, everywhere below $T_{max} \approx T_g$) 
they violate the AT stability condition and so are not physically 
relevant. 
 
Outside the regions where these ergodic solutions are allowed, we have 
to look for glassy solutions.  We do this in the next section.

\section{Glassy phases} 
 
In a glassy phase, $\hat r(1)\ne 0$ and Eq.~(\ref{MSC}) has to be 
kept, which gives, 
\begin{equation} 
  T^2 = - b^3 {\cal V}''(b) 
  \label{b(T)} 
\end{equation} 
Also, equations (\ref{Sigma}), (\ref{qtq}) and (\ref{qq0}) can be 
rewritten in the form 
\begin{eqnarray} 
  && \frac{ {\cal V}'(b) - {\cal V}'(b_0) }{ b_0 - b } = \frac{T^2}{2} 
      \frac{\sqrt{\tilde\mu}}{b} 
        \left( \frac{1}{b} + \sqrt{\tilde\mu} \right), 
  \label{b0b} \\ 
  && b_0 - b = \frac{1}{x} \left( \frac{1}{\sqrt{\tilde\mu}} - b \right). 
  \label{xbb0} 
\end{eqnarray} 
and, with $\mu_0=\mu$, (\ref{a}) and (\ref{mut}) become 
\begin{eqnarray} 
   a &=& \frac{b_0}{2} +\frac{1}{8T^2 \tmu^{3/2} 
   (1 + b_0)^{\frac{d}{2}+1}} + \frac{1}{4\sqrt{\tmu}} 
   \left( 1 + \frac{\mu}{\tmu}\right) 
\lb{eq:a2}  \\ 
\tmu &=& \mu + \frac{1}{TT_0 (1 + a)^{\frac{d}{2}+1}} 
\lb{eq:tmu2} 
\end{eqnarray} 
The above equations can be solved as follows. Eq.~(\ref{b(T)}) gives 
$b$ as a function of $T$, and then (\ref{b0b}), (\ref{eq:a2}) and 
(\ref{eq:tmu2}) can be used to find $b_0$ and $\tmu$ as functions of 
$T$ and $T_0$. Once $b_0$ and $\tmu$ are found one can calculate 
$\tilde q$ as 
\begin{eqnarray} 
 &  \tilde q = & \frac{b_0}{2} + 
  \frac{1}{8T^2 \tilde\mu^{3/2} 
     (1 + b_0)^{\frac{d}{2}+1}} + \nonumber \\ 
 & &  +\frac{1}{4\sqrt{\mu}} ( 2 + \sqrt{\frac{\mu}{\tilde\mu}}) 
  (1-\sqrt{\frac{\mu}{\tilde\mu}}) 
  \label{qtglas1} 
\end{eqnarray} 
 
As in our analysis of ergodic solutions in the preceding section, we 
will try two types of ansatz: one with $\tmu/\mu=const$ as 
$\mu\rightarrow 0$ and one with $\tmu=const$ as $\mu\rightarrow 0$ 
leading to what we call frozen-globule and glassy native phases, 
respectively.

\subsection{Frozen-globule phase} 
 
The limit where $\alpha=\tmu/\mu$ is kept constant and $\mu\rightarrow 
0$ is easily treated. Eq.~(\ref{b(T)}) stay the same, while 
Eq.~(\ref{b0b}) gives 
\begin{equation} 
 \frac{{\cal V}(b)-{\cal V}(b_0)}{b_0-b} 
 \approx\frac{T^2\sqrt{\alpha}}{2b^2}\sqrt{\mu}. 
\end{equation} 
Since $b$ is kept fixed the only solution of equation above is 
$b_0\rightarrow\infty$ as 
\begin{equation} 
  b_0 \approx \frac{\psi(b)}{\sqrt{\alpha\mu}}, 
  \label{b0glas} 
\end{equation} 
where $\psi(b)$ is a function which depends only on $b$, as $\mu$ is 
sent to $0$. Inserting (\ref{b0glas}) into (\ref{eq:tmu2}) gives 
\begin{equation} 
  \alpha = 1 + {\cal O}(\mu^{(d-2)/4}) 
\end{equation} 
and $\alpha$ stays very close to $1$, as in the ergodic random globule 
case.  Also, $\varphi$ is given by (\ref{varphi2}), while 
(\ref{qtglas1}) gives 
\begin{equation} 
 \tilde q = \frac{1}{2\sqrt{\mu}} \left[ \psi(b) 
                 + {\cal O}(\alpha-1) \right] 
  \label{qtglas2} 
\end{equation} 
which can be compared with ergodic globule result, Eq.~(\ref{qterg2}). 
Eq.~(\ref{costh}) stays the same, and one gets $\cos \theta\sim 
\alpha-1$ which goes to zero as $\mu\rightarrow 0$. There is no 
overlap with native state. Does the system freeze into some other 
configuration? To find out, we calculate overlap angles between 
configurations at time $t$ and a much later time $t'$.  As discussed 
in section \ref{sec:order_parameters} there are two ways in which the 
limit $t,t\rightarrow\infty$ can be taken, leading to $q_0\ne q$. 
 
In the first limit, the equivalent of Eq.~(\ref{costh'}) for the 
ansatz used here reads 
\begin{equation} 
   \cos \theta'_g = 
   \frac{ 
     \lim_{\lambda,t\rightarrow\infty} \langle x(s,t)x(s,\lambda t)\rangle 
    }{ 
      \sqrt{ [\lim_{t\rightarrow\infty} \langle x(s,t)^2 \rangle ]^2 } 
    } = 
   \frac{q_0}{\tilde q} 
   \label{costh'g} 
\end{equation} 
Using Eq.~(\ref{bqq0}), we can write $\cos \theta'_g =1-b_0/2\tilde 
q$, and Eqs~(\ref{qtglas2}) and (\ref{b0glas}) give $\cos \theta'_g 
\sim \alpha-1$ which goes to $0$ as $\mu\rightarrow 0$. (This behavior 
is analogous to that found in p-spin glasses.)  However, at 
not-too-long time scales (shorter than the waiting time), as in 
equation (\ref{costh''g}), the polymer is frozen: 
\begin{equation} 
   \cos \theta''_g = 
   \frac{ 
     \lim_{\tau,t\rightarrow\infty} \langle x(s,t)x(s,t+\tau)\rangle 
    }{ 
      \sqrt{ [\lim_{t\rightarrow\infty} \langle x(s,t)^2 \rangle ]^2 } 
    } = 
   \frac{q}{\tilde q} 
   \label{costh''g} 
\end{equation} 
Using Eq.~(\ref{bqq0}), we can write $\cos \theta''_g=1-b/2\tilde q$, 
and Eq.~(\ref{qtglas2}) gives $\cos \theta''_g\sim 
1-\sqrt{\mu}b/\psi(b)$, which goes to $1$ as $\mu\rightarrow 0$. 
Thus, this glassy phase has no overlap with the native state. 
 
As discussed above, there is an upper temperature limit $T_g$ 
(independent of $\beta_0$) above which this phase melts, leaving the 
system in the random-globule state.  $T_g$ can be found from 
Eqs.~(\ref{b(T)}) and (\ref{b0b}), using $b_0 \rightarrow \infty$ and 
(\ref{xbb0}) with $x \rightarrow1$.  This leads to a value $b = 
2/(\half d - 1) = {\cal O}(1)$ at the transition and $T_g = 2 (\half d 
-1)^{\half (\half d - 1)}/(\half d +1)^{\half( \half d +1)}$.  For 
$d=3$, $T_g \approx 0.535$.

\subsection{Glassy native states} 
 
We also have to study the possible glassy phases with overlap with the 
native state, i.e., with finite $\tmu$ (and, accordingly, finite $a$) 
when $\mu\rightarrow 0$.  In such a phase, as in the ergodic 
native-like states described above, the system moves only in the 
neighborhood of the native state configuration.  However, in a 
``glassy native'' state even this restricted motion is strongly 
suppressed by the complexity of the local potential energy surface, 
and a glassy phase results. 
 
As in the ergodic ansatz, the limit $\mu\rightarrow 0$ introduces no 
problems.  Eqns.~(\ref{b(T)}) and (\ref{b0b}) remain the same as in 
the frozen-globule case, while the equations for $a$ and $\tmu$, 
(\ref{eq:a2}) and (\ref{eq:tmu2}) become 
\begin{eqnarray} 
  a(\tmu) &=& \frac{b_0}{2} +\frac{1}{8T^2 \tmu^{3/2} 
    (1 + b_0)^{\frac{d}{2}+1}} + \frac{1}{4\sqrt{\tmu}} 
    \lb{eq:a3} \\ 
  \tmu(a) &=& \frac{1}{TT_0 (1 + a)^{\frac{d}{2}+1}} 
     \lb{eq:tmu3} 
\end{eqnarray} 
Again, Eq.~(\ref{b(T)}) specifies $b$ as a function of $T$, and 
(\ref{b0b}), (\ref{eq:a3}) and (\ref{eq:tmu3}) determine $b_0$ and 
$\tmu$ as functions of $T$ and $T_0$. $\tilde q$ and $\varphi$ are 
given by $\varphi,\tilde q\approx 1/(2\sqrt{\mu})$. 
 
The overlap with the native state is the largest possible: $\cos 
\theta_g=1$, as can be easily seen from Eq.~(\ref{costh}) and the 
values for $\varphi$ and $\tilde q$ we have just given. The overlap 
between two conformations at very different times also takes its 
largest possible value. From Eqs.~(\ref{costh'g}) and 
(\ref{costh''g}), knowing that $b_0$ and $b$ do not depend on $\mu$ we 
have $\cos \theta'_g =1-b_0/2\tilde q\approx 1 - b_0 
\sqrt{\mu}\rightarrow 1$ and $\cos \theta''_g =1-b/2\tilde q\approx 
1-b\sqrt{\mu}\rightarrow 1$. Thus, the polymer is frozen almost 
everywhere into the native conformation. However, the freezing is not 
total, since $a$ in (\ref{eq:a3}) is not zero. Furthermore, there is 
aging in the system, since $x$ in Eq.~(\ref{xbb0}) is not equal to 1. 
 
We turn now to the solution of the equations (\ref{b(T)}), 
(\ref{b0b}), (\ref{eq:a3}) and (\ref{eq:tmu3}).  As for the 
corresponding ergodic phases we have to resort to numerical solution; 
here we describe the analysis.  The working strategy is similar to the 
one presented in subsection IX.B; the goal is to find $\beta_0$ as 
function of $\tmu$ for fixed $T$ since, as in the ergodic native case, 
extrema of the function $\beta_0(\tmu,T)$ govern the phase boundaries. 
 
The procedure for finding value of the function $T_0(\tmu,T)$ is as 
follows. Eq~(\ref{b(T)}) determines $b$ as a function of $T$, to be 
referred to as $b(T)$. Once $b(T)$ is found from (\ref{b(T)}) it is 
inserted into Eq.~(\ref{b0b}), which determines $b_0(T,\tmu)$. The 
value found for $b_0$ is inserted into Eq.~(\ref{eq:a3}) to find $a$, 
and finally $\beta_0 =1/T_0$ is calculated from 
Eq.~(\ref{eq:tmu3}). Thus, at each temperature for which glassy 
solutions are possible, we can construct a graph of $\beta_0(\tmu)$, 
as we did for ergodic solutions in Fig.~2.  We have used Mathematica 
to do these calculations.  We can use these curves, together with the 
ergodic ones previously analyzed, to identify the possible states of 
the system at a given temperature and $\beta_0$ (Fig.~4). The 
procedure is fairly simple.  At any given $\tmu$, only one of the 
solutions is physical: In the region $\tmu_{min}<\tmu<\tmu_{AT}$ one 
has to follow the glassy $\beta_0(\tmu)$ curve, while outside it one 
follows the ergodic one.  In Fig.~4 the physical solution is indicated 
as the thick dashed curve.  One then looks for solutions as 
intersections of this curve with a horizontal line at a given value of 
$\beta_0$, as done previously (e.g., as in Fig.~2, panel (a)) within 
the ergodic ansatz.

In Fig.~4 this procedure is shown for several different values of $T$. 
In the first panel, $T$ lies just a little below $T_{max}$ (as in 
Panel (b) of Fig.~2).  Suppose we start in the ergodic native phase at 
large $\beta_0$ and then lower $\beta_0$.  (In Fig.~5, this would 
correspond to moving along a horizontal line (constant $T$) slightly 
above point B in panel (b) or (c)).  We can lower $\beta_0$ all the 
way down to $\beta_0^{\rm min}$ without encountering an AT 
instability.  So, just as in the ergodic analysis of section X.B, 
beyond $\beta_0^{\rm min}$ the ergodic native phase melts into the 
random-globule phase. 
 
In the same panel we can also analyze the what happens to the unstable 
stationary state in the same range of $\beta_0$ for this temperature. 
At very large $\beta_0$ we have an ergodic solution, but as we lower 
$\beta_0$ we pass through a range of $\tmu$, between $\tmu_{min}$ and 
$\tmu_{AT}$, where the ergodic solution is unstable against 
glassiness.  In this region we must follow the glassy curve instead of 
the ergodic one.  We interpret this glassy solution in the following 
way: The free energy landscape near the US maximum becomes rough in 
this range of values of $\beta_0$ (at this temperature), the same way 
the free energy landscape near the minimum corresponding to a 
thermodynamic phase becomes rough in a glassy state.  We call it a 
``glassy US state''. 
 
The next panel is for a slightly lower temperature (but still above 
$T_g$).  Here, as we lower $\beta_0$ in the ergodic-native phase, we 
reach an AT instability before we get all the way down to the minimum 
on the ergodic curve.  (In Fig.~5, this would correspond to moving on a 
line of constant $T$, meeting the AT line somewhere between points A 
and B in panel (b) or (c)).  Furthermore, the only available glassy solution 
for $\tmu < \tmu_{AT}$ is one with negative $\partial \beta_0/\partial \tmu$, 
that is, it corresponds to the kind of glassy US state discussed 
above.  As this is not a stable phase, we conclude that for this $T$, 
the minimum value of $\beta_0$ lies at this AT line, and beyond it 
there is no stable native-like state.  We can follow the glassy US 
state back up to larger $\beta_0$, seeing that we eventually cross 
over to a normal (non-glassy) transition state. 
 
In the last panel, the temperature is lowered a bit more (below 
$T_g$).  Again, starting in the ergodic native phase at large 
$\beta_0$ and lowering $\beta_0$, we encounter an AT instability and a 
glassy solution appears. (Equivalently, in Fig.~5 one moves on a 
horizontal line somewhere below $T_g$ until meeting the AT line for 
the first time.)  For smaller $\beta_0$, we switch to the glassy 
curve, which has positive $\partial \beta_0/\partial \tmu$, describing 
a glassy native phase.  We can follow this curve down to its minimum 
$\beta_0$, beyond which no phases correlated with the native phase 
exist.  But, of course, following it back up toward large $\beta_0$ on 
the unstable branch, we can identify the glassy US state between 
the phase correlated with the native state and the one uncorrelated 
with it.  (Above $T_g$, the latter is the random-globule phase; below 
it, it is the frozen-globule phase.)

\section{Features of the phase diagram} 
 
The phase structure implied by this simple model is not so simple. 
Fig.~5 shows the phase diagram constructed from the above analysis. 
For clarity, we show in the top panel only the solutions that 
correspond to stable phases.  The second panel shows the details in 
the region where the ergodic-native, glassy-native, random-globule and 
frozen globule states come together (or nearly so).  The third panel 
shows the regions where ergodic and glassy US states are found. 
 
There are six distinct regions in the phase diagram.  In region I 
(high $T$, small $\beta_0$), the only stable phase is the random 
globule.  In region II (small $\beta_0$, $T <T_g$) it undergoes a 
glass transition to the frozen-globule phase.  The properties of the 
system in this part of the phase diagram are the same as in the 
completely random heteropolymer model of paper I; the bias of the 
interactions toward a native state does not have any effect until a 
(temperature-dependent) threshold $\beta_0^c(T)$ is reached.  This 
threshold is marked on the diagram by the lines separating region I 
from regions III and V and region II from region VI. 
 
To help thinking about these phases, we offer the schematic free 
energy-surface pictures of Fig.~6.  They show how we imagine the free 
energy varies as a function of the native-state overlap coordinate 
$\varphi$.  Fig.~6A depicts this cross-section through the free-energy 
surface in region I, where there is a single smooth minimum around 
$\varphi = 0$, representing the random-globule phase.  Fig.~6B shows 
what happens in region II, where this phase is replaced by the 
frozen-globule phase.  We represent this by drawing the free energy 
surface with many local minima.  Fig.~6C shows what happens in the 
middle of region III, where there are two (smooth) minima, the new one 
corresponding to the ergodic native phase.  (It will lie above or 
below that at $\varphi = 0$ according to whether we are above or below 
a first-order transition that we expect to occur at a temperature $T_1 
<T_n$, see below.)  In Fig.~6D we depict the situation in region IV, 
where both the $\varphi \approx 0$ region and the region around the 
maximum become rough.  Fig.~6E represents region V, with the native 
valley and the maximum rough, but the region near $\varphi = 0$ still 
smooth.  In Fig.~6F (region VI), that, too, becomes rough. 
 
An important feature is the fact that the random-globule and 
frozen-globule phases remain (dynamically) stable in their respective 
temperature ranges for all $\beta_0$.  Thus the horizontal line 
separating region I from region II continues across the diagram, 
separating region V from region VI and region II from region IV. 
 
In each of regions II, IV, V and VI (Fig.~6, panels (b), (d), (e), 
(f)) there are two stable states.  Thus, regions I and III are 
separated by the line $T_n(\beta_0)$, below or to the right of which, 
in addition to the random-globule state, an ergodic native 
state exists and is stable.  In region IV, the frozen-globule and this 
ergodic native state are both stable.  In region V, the random globule 
state and a glassy native state are stable, while in region VI the 
stable states are the frozen globule and the glassy native phase. 
 
The diagram shows some interesting fine structure in the neighborhood 
of the region where regions I, III, and V meet (second panel, point 
A).  In particular, below point A, the boundary between regions III 
and V (i.e., between the ergodic-native and glassy-native state) is an 
AT line.  It comes about as can be seen in the last panel of Fig.~4, 
where, following the ergodic phase down from high $\beta_0$ and 
$\tmu$, we reach $\tmu_{AT}$ and thereafter have to switch to the 
glassy solution. 
 
On the other hand, the boundary $T_n(\beta_0)$ above point B is 
reached as in the first panel of Fig.~4: one can come all the way down 
to the minimum value of $\beta_0$ for the ergodic solution before 
reaching $\tmu_{AT}$.  The full line continuing upwards and to the 
right of point B is an AT line which goes over into an $x=1$ line 
at its maximum, $T_{max}$.  Below this line US solutions become 
glassy. 
 
Between points A and B, the boundary is marked by reaching $\tmu_{AT}$ 
in the way shown in the second panel of Fig.~4: There is no stable 
ergodic native solution to the left of this line, since, upon lowering 
$\beta_0$ below $\beta_0^{\rm min}$ in Panel (b), the solution with 
$\partial \beta_0/\partial \tmu>0$ is lost.  Thus in this region the 
line AB is an AT line for both the native phase and the US solutions. 
The dashed line marks the boundary found if one ignores the AT line 
(i.e., it is a portion of the boundary found using the ergodic ansatz 
and shown in Fig.~3). 
 
Everywhere below the AT line (both the portion AB and its extension 
upward to the right) is the region where the US solutions are glassy, 
as shown in the last panel of Fig.~5.  At a given $\beta_0$, these 
features all have an onset at a temperature between $T_g$ and 
$T_{max}$.  As can be seen in Fig.~1, this is a very small temperature 
range.  This is also the reason why there is fine structure, as shown 
in both the second and the third panels of Fig.~5, in such a small 
temperature range in the phase diagrams. 
 
As remarked above, we have not done an equilibrium calculation, but we 
expect that the first-order transition temperature $T_1(\beta_0)$ 
where the free energies of the random-globule and ergodic-native 
phases are equal will also rise with $\beta_0$.  For large $\beta_0$, 
we expect $T_1$, like $T_n$, to be proportional to $\beta_0$ (but $T_1 
< T_n$, of course). 
 
Thus, at fairly large $\beta_0$, we expect the following sequence of 
stable states as we lower $T$ from a high value.  Initially, only the 
random-globule state is stable.  Then, below $T_n$, the ergodic native 
state is also stable, and below $T_1$ it becomes the lowest-free 
energy state.  Going further down in $T$, we cross the boundary (last 
panel in Fig.~5) where the US state between the ergodic native and 
random-globule states becomes glassy (i.e., acquires a rough local 
free energy landscape).  Very soon thereafter, we cross $T_g$, where 
the random-globule state undergoes glassy freezing.  Continuing, we 
reach a temperature where the ergodic native state undergoes glassy 
freezing.  Finally, we reach the stability limit of this glassy-native 
phase, leaving the system with nowhere to go but the frozen-globule 
state. 
 
What lessons are there in these findings for protein folding?  We 
start from the assumption that the initial state in the folding 
process is uncorrelated with the native state (i.e., in Fig.~6 we 
start in a local minimum at $\varphi\approx 0$).  Folding requires the 
system to find its way to the (ergodic) native state. 
 
One feature that is evident is that such a path in configuration space 
always requires an uphill free-energy step.  This is because either 
the random-globule or frozen-globule phases is always locally stable. 
 
If we stick to our mean-field dynamical picture, where barriers are 
infinite, folding is, strictly speaking, impossible.  In dynamical 
terms the ``infinite'' barriers translate into the fact that the 
equations which govern the motion of order parameters, $\varphi$ 
included, have basins of attraction corresponding to the plots shown 
in Fig.~6. For example, starting from $\varphi$ somewhere close to $0$ 
and given a free energy profile like that in Fig.~6C, dynamical 
equations will never carry $\varphi$ to the large value describing the 
native state. On the contrary, $\varphi$ will approach $0$ as time 
goes on. 
 
But, if we relax this assumption and imagine finite barriers 
(associated with local nucleation of a native 
phase\cite{BW90,TakWol97,PTW01}), we may ask (informally) when 
activated motion over the barrier to nucleation is least hindered.  We 
argue that the free energy landscape features present globally in our 
mean-field picture will also be relevant locally: when our calculation 
here finds a glassy US state, we expect that free energy surface near 
the true transition state will also be rough.  Thus, from the 
preceding description of the phases and the transition states between 
them, we can see that folding should be easiest for large $\beta_0$ in 
a window between $T_1(\beta_0)$ and the upper boundary of the region 
where the US states become glassy (and passage across the transition 
region is kinetically impeded by the tortuous nature of the local free 
energy landscape).  The latter boundary lies, in turn, just barely 
above $T_g$, where the landscape in the (large) portion of the 
configuration space uncorrelated with the native state also becomes 
rough, further impeding escape from it.  At still lower temperatures, 
things become even worse, first with the onset of glassiness in the 
native-like region of configuration space itself and finally with the 
disappearance of native-like solutions.  But these features probably 
have minor consequences, since folding will already have been so 
strongly impeded by the effects (with onset near $T_g$) that tend to 
confine it in a region of configuration space uncorrelated with the 
native state.

\section{Discussion} 
 
We have introduced what we might call a generic model for a protein, 
based on what seems to us to be the simplest way to incorporate a 
tendency to form a native state in an otherwise random heteropolymer 
model.  To make it possible to calculate typical properties, we follow 
previous authors \cite{PGT1,RS,PGT2,WS} and do not specify a 
particular native state, but rather an ensemble of them, constrained 
only by chain-entropic constraints and confinement to the appropriate 
volume.  This ensemble is characterized by the selection temperature 
$T_0$.  Our model differs from previous ones in that they are based on 
random-sequence heteropolymers, while we start from a model 
\cite{SG1,SG2} in which each monomer-monomer interaction is an 
independent random variable. 
 
While it might be argued that random-sequence models are more relevant 
to proteins, they approach the model we consider here in the limit 
where the number of monomer types becomes large.  Thus, what we find 
out about our model may be relevant to proteins (with 20 different 
amino acids).  Of course, it is also important to study what happens 
away from the large-monomer-type limit; our analysis here can help in 
solving that more difficult problem. 
 
Furthermore, naively, one might assume that by adjusting $N(N-1)/2$ 
parameters one could imprint a native state more strongly than for 
models with only $N$ parameters. Our model shows that this is not 
necessarily true.  Parts of the phase diagram are glassy, even for 
very low selection temperature $T_0$, when the native state should be 
strongly imprinted into the model. 
 
Instead of the quadratic confinement term $\mu x(s,t)^2$ one could add 
three-body terms, which are commonly used to fix the globule 
density. It would be interesting to extend the analysis presented here 
to such models.  Also, in our treatment, translational invariance 
within the globule is put in by hand.  Keeping three-body terms would 
lead to automatic translational invariance.  We have seen that if 
translational averaging is omitted (see paper I) then the equations 
become coupled in the $k$ variable and are thus a lot harder to solve. 
 
Within out model, we have made just two approximations: the Gaussian 
variational ansatz of section VI, and the assumption of 1-step 
ergodicity breaking (analogous to 1-step replica symmetry breaking in 
the replica approach).  Otherwise the solution is complete and exact 
to the accuracy we were able to achieve numerically. 
 
Our most important result is the existence of the various different 
phases at large $\Bt/T_0$, where the interactions that favor a native 
state are strong.  While it is natural to anticipate that the 
native-like configurations will be thermally disrupted above a 
temperature of order $\Bt^2/T_0$, it is not so obvious that at low 
temperatures there will be other impediments to efficient folding.  We 
identify two of these: 
 
(1) The frozen-globule state, which is uncorrelated with the native 
state, always exists below $T_g$, no matter how big $\beta_0$ is. 
This means that in a large part of configuration space, the system may 
be trapped in a rough energy landscape and never (in MFT) get to the 
native-state region where it can fold rapidly. Furthermore, in almost 
the same temperature, range, we expect that the energy landscape is 
also rough around the transition region on the way to the 
correctly-folded state, further impeding the folding process.  Thus, 
while lower temperature favors well-folded over random-globule-like 
configurations energetically, the rough energy landscape of the glassy 
phase will hinder correct folding.  Our conclusion here is consistent 
with that of Goldstein {\em et al.}\cite{GLW92}, who found (albeit in 
a different kind of model) that a large $T_n/T_g$ (or $T_1/T_g$) ratio 
favors folding. 
 
(2) At even lower temperatures, the native state itself is 
unstable against a glass transition where it splits into a large 
number of substates.  Transitions between these substates are 
blocked by high barriers (infinite, in MFT).  A phase of this kind 
was found earlier by Bryngelson and Wolynes in a phenomenological 
model\cite{BW87}.  It is tempting to associate the substates with 
the glassy conformations observed at temperatures below ~200 K in 
myoglobin \cite{Ibenetal}. 
 
Of course, MFT is an approximation.  The escape from the tortuous part 
of the energy landscape to the smooth region will not take forever, 
nor will transitions between low-$T$ substates.  Nevertheless, MFT 
does indicate when we can expect relaxational dynamics, including 
folding, to be slow or fast, as well as give us some insight into the 
physics behind these differences. 
 
Our analysis here is a purely dynamical one. We do not compute 
equilibrium partition functions.  A complete analysis would include 
such calculations, but we defer them to future work.  Nevertheless, 
the purely dynamical analysis can reveal important properties of the 
system that cannot be seen in an equilibrium analysis.  For example, 
it has been know for a long time that for a large class of models --- 
namely, those which have a glass transition where $x \rightarrow 1$ 
--- the dynamic and static glass transition temperatures are different 
\cite{CK1,CS,CHS}.  This is expected to be the case for the transition 
at $T_g$ in our model: The equilibrium glass transition temperature is 
lower than the dynamical one.  Thus, in a temperature range just below 
the dynamical $T_g$, the equilibrium analysis does not reveal the slow 
dynamics (accompanied by aging) that we are able to identify and 
analyze here. 
 
In other glassy models for which it has been possible to do a more
complete analysis \cite{CKD,CD,CK1,CS,CHS}, the static and dynamic
transitions coincide when they occur as a result of an
Almeida-Thouless instability (the marginal stability condition,
Eqn~\ref{MSC}) \cite{AT}.  This is the case here at the phase boundary
where the native-like state becomes glassy.
 
Gillin and Sherrington \cite{GS} and Gillin {\em et al.} \cite{GNS}
have been able to analyze both the statics and the dynamics of several
classes of mean-field spin glass models with a competition between
glassy and ferromagnetic states (see also ref.~\cite{HSN} for a
special case).  Some features of the phase diagram of our model that
we have been able to discover so far are also seen in their models.
 
Gillin {\em et al.} also studied full (as well as 1-step) replica
symmetry breaking solutions, which we have not.  In some of their
models, the counterpart of our glassy native phase undergoes full RSB
at low temperatures, and the counterpart of our II-VI boundary becomes
vertical. It is possible in our model as well that, in particular
regions of the phase diagram, our 1-step solutions are not stable and
full RSB is necessary. More generally, it will be an interesting
problem to try to explore what kinds and degrees of universality there
are in the phase structures of various systems where the glassiness
induced by disorder competes with some kind of order analogous to the
native state in our problem or the ferromagnetic state in theirs.

\begin{acknowledgements} 
It is a pleasure to thank Silvio Franz for discussions leading to our 
formulation of this problem. 
\end{acknowledgements} 
 
\appendix 
 
\section{correlation function $G_{10}^{ss'}$ } 
 
Here we derive Eq.~(\ref{G10a}). Inserting Eq.~(\ref{Phis1}) for 
the superfield  into (\ref{G10}) gives 
\begin{eqnarray} 
  G_{10}^{ss'} 
    & = & \langle x(s,t_1) x_0(s') \rangle 
        + \langle \bar\eta(s,t_1) x_0(s') \rangle \theta_1 + \nonumber \\ 
    &   & + \bar\theta_1 \langle \eta(s,t_1) x_0(s') \rangle 
          + \bar\theta_1\theta_1 \langle \tilde x(s,t_1) x_0(s') 
          \rangle . 
  \label{G10b} 
\end{eqnarray} 
One can show that the action of the dynamical generating 
functional $F_{dyn}$ (see Eq.~\ref{Fdyn}) is invariant under the 
infinitesimal transformation (BRS symmetry) 
\begin{equation} 
  \delta \Phi(s,t,\theta,\bar\theta) = \epsilon 
    \frac{\partial}{\partial\bar\theta} \Phi(s,t,\theta,\bar\theta) 
  \label{deltaPhi}. 
\end{equation} 
This follows in two steps. First one notices that for any function 
$f$ 
\begin{equation} 
  \delta\int d\bar\theta f(\Phi(\bar\theta)) = 
    \epsilon \int d\bar\theta \frac{\partial}{\partial\bar\theta} 
       f(\Phi(\bar\theta)) = 0, 
\end{equation} 
due to the identity $\int d\bar\theta 
\frac{\partial}{\partial\bar\theta} = 0$. This means that any term 
involving a local function of $\Phi$ (i.e., not containing 
derivatives over $\theta$ and $\bar\theta$), e.g., $S_2[\Phi,x_0]$ 
(see Eq.~\ref{eq:S2}), is invariant under the transformation 
(\ref{deltaPhi}). $S_0[x_0]$ is trivially invariant since it does 
not contain the superfield $\Phi$.  (The same reasoning holds for 
a transformation like (\ref{deltaPhi}) with a derivative with 
respect to $\theta$ instead of with respect to $\bar\theta$.)  The 
only term left is the $S_1[\Phi]$ (i.e., the part of the action 
quadratic in the superfield) and it is straightforward to see that 
this term is also invariant under (\ref{deltaPhi}) (though not 
under the transformation involving the derivative with respect to 
$\theta$). 
 
The fact that the action is invariant under (\ref{deltaPhi}) 
implies the Ward identity 
\begin{equation} 
  \frac{\partial}{\partial\bar\theta_1} G_{12}^{ss'} = 0 
  \label{ward1} 
\end{equation} 
which gives $0=\langle \eta x_0 \rangle + \theta_1 \langle \tilde 
x x_0 \rangle$ (we have suppressed the arguments of the fields to 
simplify the notation).  This implies that separately one has 
\begin{equation} 
  \langle \eta x_0 \rangle=0, \ \ \langle \tilde x x_0 \rangle=0 
  \label{ward2} 
\end{equation} 
Inserting (\ref{ward2}) into (\ref{G10b}) gives the desired 
result, Eq.~(\ref{G10a}). 
 
\section{Details on the GVA and how to improve it} 
 
Here we give more background on the use of Eq.~(\ref{dFdG}).  In 
the dynamical calculation, the fields are complex and contain 
Grassmann variables; thus, $F_{dyn}$ is not a real number. This 
means that any interpretation of Eq.~(\ref{dFdG}) as an extremum 
condition for $F_{dyn}$ has to be given up. Nevertheless, we can 
still make some sense of the GVA as the first step in a systematic 
approximation scheme. 
 
Formally, one starts from Eq.~(\ref{Fdyn}), which we rewrite in the 
shorter form 
\begin{equation} 
   e^{-F_{dyn}}=\int D\Psi e^{ -S[\Psi] }. 
  \label{Fdyn3} 
\end{equation} 
where $\Psi$ stands for the pair $(x_0,\Phi)$ and, likewise, 
$D\Psi$ for $Dx_0D\Phi$. One can express $F_{dyn}$ in a slightly 
different form 
\begin{equation} 
  e^{-F_{dyn}} = \langle e^{-(S-S_{var})} \rangle_{var} e^{-F_{var}} 
 \label{Fdyn4} 
\end{equation} 
where 
\begin{equation} 
  e^{-F_{var}}=\int D\Psi e^{-S_{var}} 
\end{equation} 
and 
\begin{equation} 
 \langle (...) \rangle_{var}= 
   \frac{ 
     \int D\Psi ( ... ) e^{-S_{var}} 
    }{ 
     \int D\Psi e^{-S_{var}} 
    } 
\end{equation} 
In a static calculation one would proceed with the inequality 
\begin{equation} 
   e^{-F} \ge e^{  - \langle (S-S_{var}) \rangle_{var} } 
              e^{ -F_{var} } 
   \label{Fdyn5} 
\end{equation} 
to conclude that 
\begin{equation} 
   F \le \langle S-S_{var} \rangle_{var} + F_{var} 
   \label{Fdyn6}. 
\end{equation} 
Thus, in a static calculation, the variationally-obtained $F$ 
gives an upper bound on the true $F$. What is allowed to vary is 
the form of $S_{var}$, most often, the parameters describing it. 
(In the GVA, $S_{var}$ is specified by $G_{12}^{ss'}$ and 
$G_{10}^{ss'}$.) 
 
In the dynamical problem we follow another route, starting exactly 
at the problematic step, Eq.~(\ref{Fdyn5}), along the lines of 
ref.~\cite{Kleinert}.  Instead of the inequality (\ref{Fdyn6}) we 
use Eq.~(\ref{Fdyn4}) in a slightly modified form 
\begin{equation} 
  F_{dyn} = F_{var} 
            - \ln \langle e^{-\Delta S} \rangle_{var} 
  \label{Fdyn7}, 
\end{equation} 
where $\Delta S=S-S_{var}$.  Applying  a cumulant expansion 
\begin{eqnarray} 
  \langle {\rm exp}(-\Delta S)\rangle_{var} & & = 
    {\rm exp}\left[ -\langle \Delta S \rangle_{var} + \right. \nonumber \\ 
  & &   \left. + \frac{1}{2} \left( 
       \langle \Delta S^2 \rangle_{var} 
       - \langle \Delta S \rangle_{var}^2 
      \right) 
    + \cdots \right], 
  \label{Fdyn8} 
\end{eqnarray} 
one gets 
\begin{equation} 
  F_{dyn} = F_{var} + \langle S-S_{var} \rangle_{var}  + \Delta F 
\end{equation} 
where $\Delta F$ contains second and higher order corrections in 
$\Delta S$.  In any approximation made by keeping a finite number 
of terms in (\ref{Fdyn8}) (the simplest being to set $\Delta 
F=0$), $F_{dyn}$ depends on $G_{12}^{ss'}$. To minimize this 
dependence, we chose $G_{12}^{ss'}$ so that the derivative of the 
approximate form for $F_{dyn}$ with respect to $G_{12}^{ss'}$ 
vanishes.  This gives Eq.~(\ref{dFdG}). Furthermore, if all terms 
in $\Delta F$ are kept, this procedure, by construction, formally 
gives back the exact $F_{dyn}$. 
 
The meaning of minimizing the dependence with respect to 
quantities involving Grassmann numbers may seem obscure, but we 
note that we are using the SUSY representation only for 
compactness. The entire GVA calculation could have been presented 
equivalently without any Grassmann variables, with no change in 
meaning or result. Thus, we are really only minimizing the 
dependence on parameters of physically well-defined correlation 
and response functions. 
 

\newpage 
 
\begin{figure} 
\epsfxsize=8cm 
\epsfbox{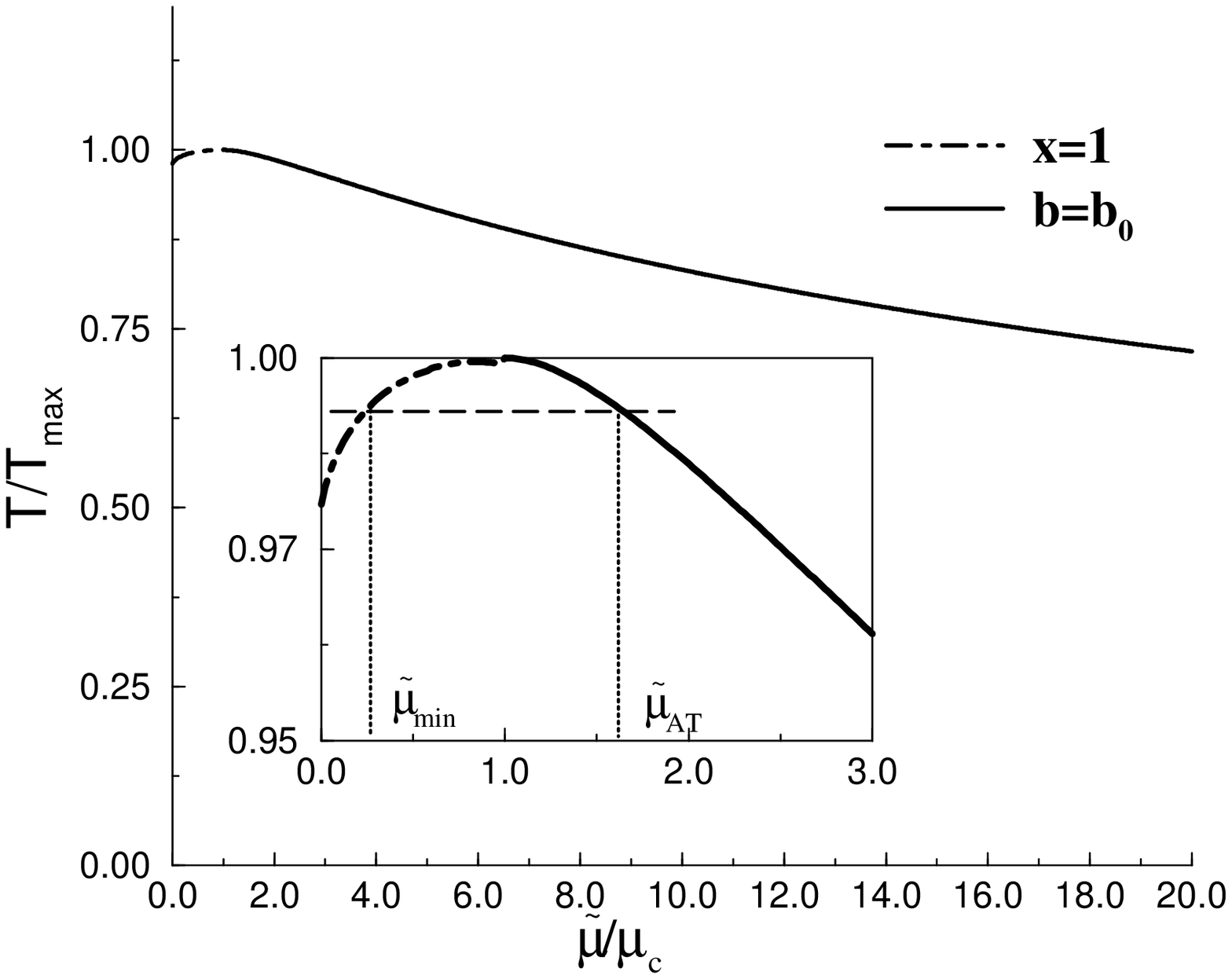} 
\caption{Boundary of the glassy phase in the $(\tmu, T)$ plane. 
The boundary is same as in the case of the random-heteropolymer 
model from paper I, except that native-state correlations lead to 
the replacement of $\mu$ by $\tilde\mu$.  We have used parameters 
$d=3$ and $\sigma=1$. $T_{max}$ is the maximum $T$ for which 
Eq.~(\ref{MSC}) has a solution (see Paper I for further details). 
$\mu_c$ is the value of $\tmu$ where $T(\tilde\mu)$ attains this 
maximum.  The solid part of the boundary is an AT line, and the 
dash-dotted part marks a transition where $x \rightarrow 1$. 
Approaching the AT line from below, $b-b_0 \rightarrow 0$, while 
$x$ remains strictly less than $1$. Approaching the $x=1$ line 
from below, $x \rightarrow 1$ smoothly, while $b-b_0$ is 
discontinuous there. Above both lines, $b=b_0$ and $x$ is 
undetermined (any $x \neq 0$ solves (74)-(80), and no physical 
quantity depends on it). The same holds for all figures where 
these lines appear.} 
\end{figure}

\begin{figure} 
\epsfxsize=9cm 
\epsfbox{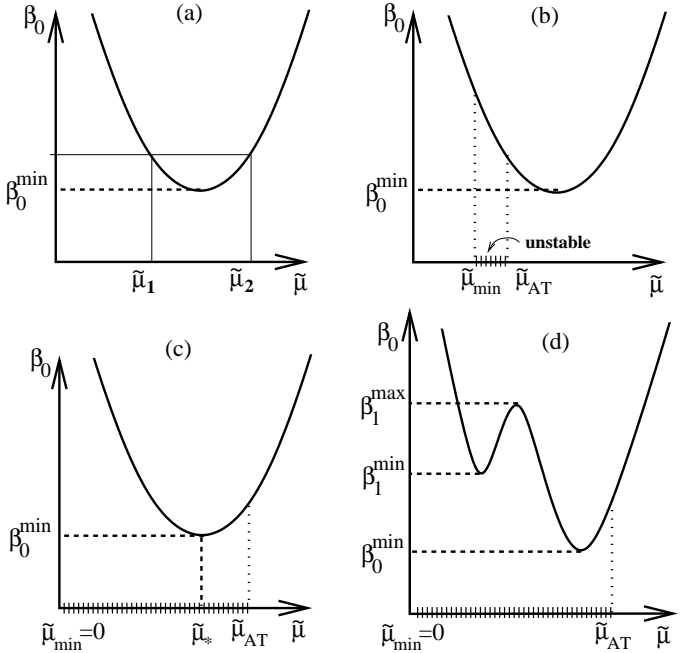} 
\caption{Analysis of ergodic-native solutions: $\beta_0(\tmu,T)$ as 
function of $\tmu$ for four values of $T$ (see text for explanation). 
Panel (d) shows the existence of the two extra solutions (one stable, 
the other unstable) in the range $[\beta_1^{\rm min}, \beta_1^{\rm 
max}]$.  } 
\end{figure}

\begin{figure} 
\epsfxsize=8cm 
\epsfbox{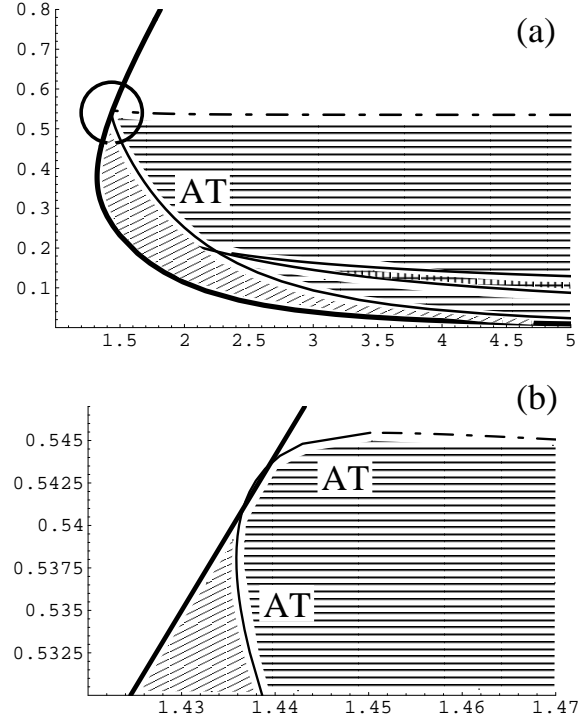} 
\caption{Regions of existence and stability of ergodic-native solutions. 
Panel (a): Ergodic-native-phase and US solutions exist everywhere to 
the right of the thick solid curve.  The ergodic native phase is 
stable against glassiness everywhere there except in the diagonally 
cross-hatched region. The US states are also unstable against 
glassiness there, and additionally in the horizontally cross-hatched 
region.  The vertical cross-hatching marks the region where the extra 
phase seen in Panel (d) of Fig.~2 is found.  (This phase is never 
stable against glassiness.)  Panel (b): Enlargement of the circled 
region in Panel (a). The AT line is tangent to the ergodic phase 
boundary (thick line).  At its maximum, at $T_{max}$, it becomes an 
$x=1$ line (dashed-dotted line, see also Fig.~1).  Lowering $T$ from 
the white region into the horizontally cross-hatched region results in 
two different types of transitions depending on whether one crosses 
the AT or the $x=1$ line.  In both cases the US state becomes glassy. 
} 
 
\end{figure}

\begin{figure} 
\epsfxsize=8cm 
\epsfbox{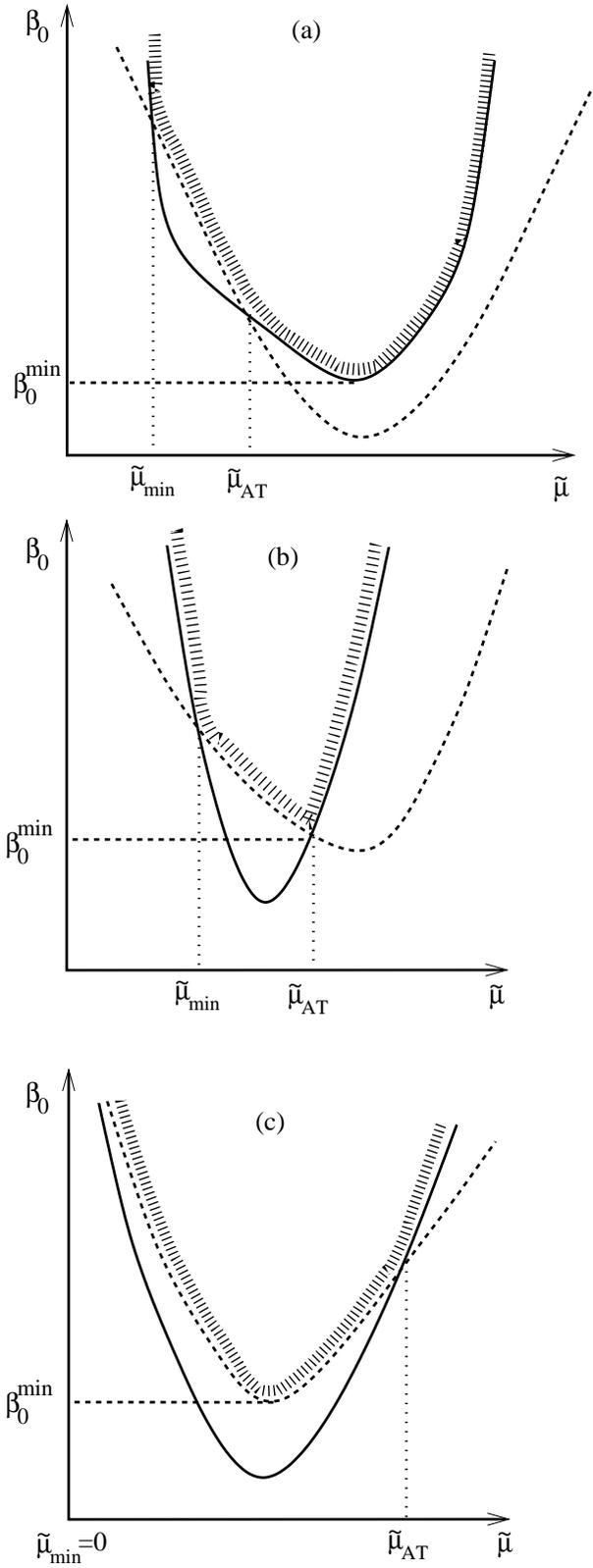} 
\caption{Analysis of glassy-native solutions: $\beta_0(\tmu,T)$ as a 
function of $\tmu$ for three fixed values of $T$ (see text for 
explanation).  Full lines: $\beta_0(\tmu,T)$ calculated within the 
ergodic ansatz (as in Fig.~2).  Dashed lines: $\beta_0(\tmu,T)$ 
calculated with the glassy ansatz.  The actual curves vary with $\tmu$ 
in a way that is difficult to plot in a useful way, so here we have 
distorted them in such a way as to make their qualitative form (number 
and ordering of maxima and minima) evident.  When the two curves cross 
at $\tmu=\tmu_{AT}$, one has to change from the ergodic to the glassy 
solution (when approaching from $\tmu=\infty$). Similarly, when 
$\tmu=\tmu_{min}$ one has to go back to the ergodic solution.  The 
thick dashed line indicates the physically relevant states (both 
stable phases and US states).  } 
\end{figure}

\begin{figure} 
\epsfxsize=9cm 
\epsfbox{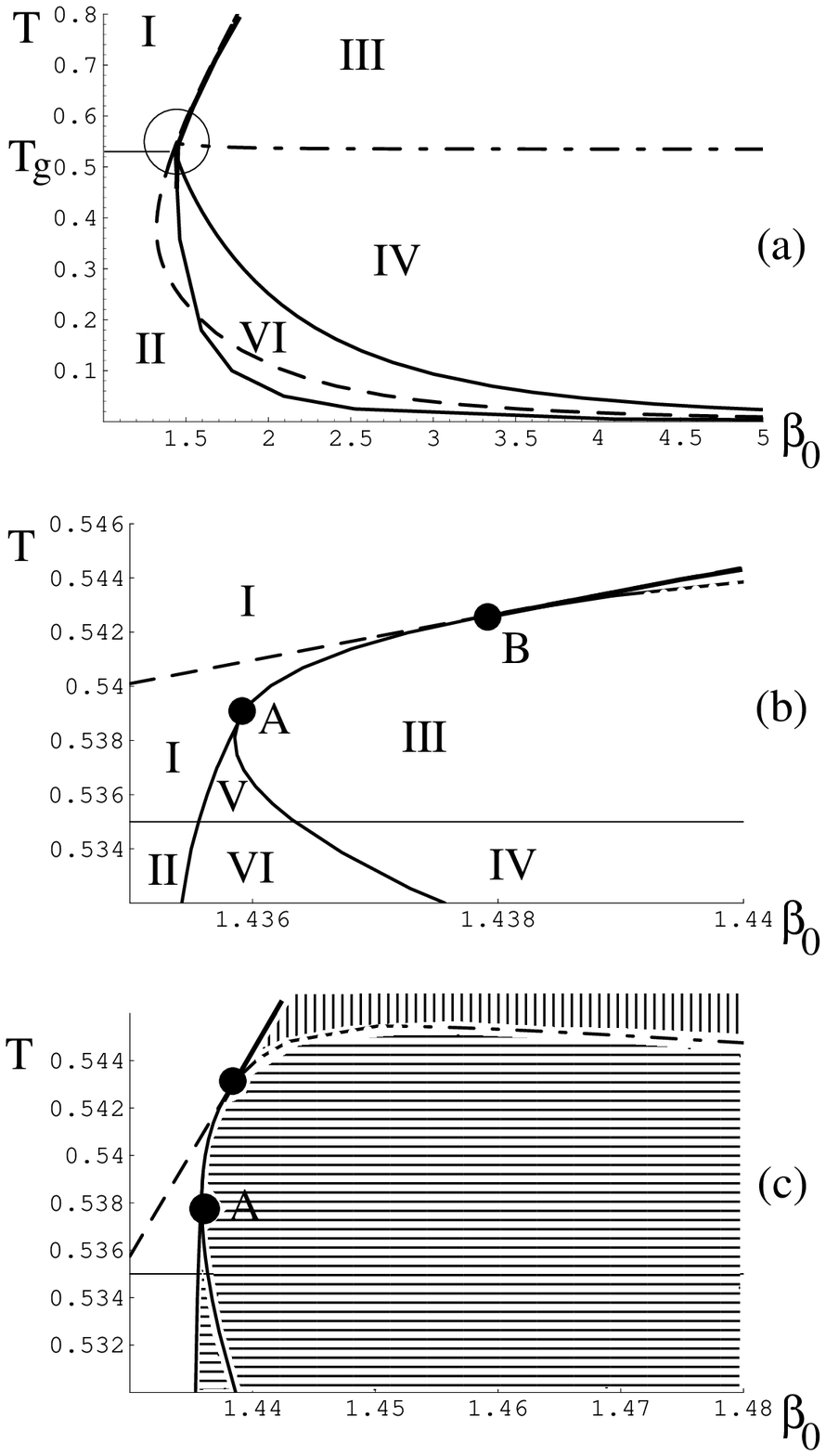} 
\caption{The final phase diagram. 
Panel (a): Stable phases. 
 Region I: Random globule is the only stable phase. 
 Region II: Frozen globule is the only stable phase. 
 Region III: Ergodic native and random globule phases stable. 
 Region IV: Ergodic native and frozen globule phases stable. 
 Region V (only visible in panel (b)): Glassy 
 native and random globule stable. 
 Region VI: Glassy native and frozen globule stable.  The dashed line 
 marks the boundary of the (unphysical) ergodic native state from 
 Fig.~3, to emphasize that the phase boundary of the glassy native 
 state (solid) does not coincide with it. 
Panel (b): Enlargement showing structure in the region near $T = 
 T_{max} \approx T_g$ and $\beta_0 = 1.45$ (including region V).  Below 
 point B the boundary of region III is given by the AT line.  Above 
 point B the boundary is the ergodic-native stability limit (the 
 uppermost line in Panel (a)).  The continuation of the AT lie is shown 
 as a dotted line (which turns into dash-dot $x=1$ line). 
Panel (c): US states are ergodic in the vertically hatched region, 
 glassy in the horizontally hatched region.  The boundary above and to 
 the right of point A is an AT line. 
 Beyond the region shown the $x=1$ line falls 
 off monotonically, and for $\beta_0 \rightarrow \infty$ it approaches 
 $T_g$.  Below point A, the small-$\beta_0$ boundary coincides with the 
 line between regions II and VI in panels (a) and (b).  } 
\end{figure}

\begin{figure} 
\epsfxsize=8cm 
\epsfbox{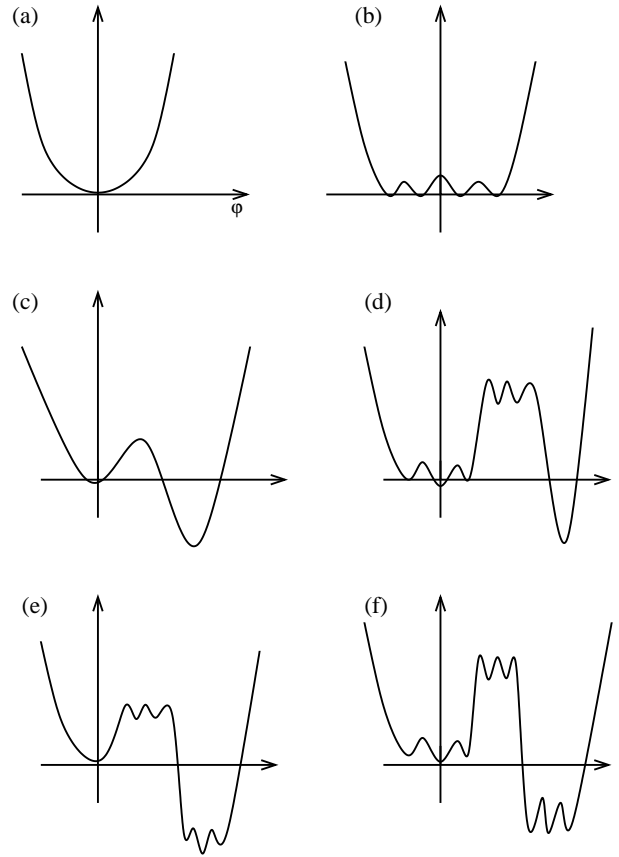} 
\caption{Schematic free energy surfaces in different regions of the 
phase diagram.  (See text for explanation.)  (a): region I.  (b): 
Region II.  (c): Region III.  (d): Region IV.  (e): Region V.  (f): 
Region VI.  } 
\end{figure}

\end{multicols} 
 
\end{document}